\documentclass[vecphys]{svmult}
\usepackage{makeidx}
\usepackage{color}
\usepackage{graphicx}
\usepackage{amsmath,amssymb}
\usepackage{multicol}
\usepackage[bottom]{footmisc}

\newcommand{\secref}[1]{\hbox{\S\,\ref{#1}}}
\newcommand{\exref}[1]{(\ref{#1})}
\newcommand{\exsref}[2]{(\ref{#1})-(\ref{#2})}
\newcommand{\exsandref}[2]{(\ref{#1}) and (\ref{#2})}

\newcommand{\eeqref}[1]{Eq.\ (\ref{#1})}
\newcommand{\eqsref}[2]{Eqs.\ (\ref{#1})-(\ref{#2})}
\newcommand{\eqsandref}[2]{Eqs.\ (\ref{#1}) and (\ref{#2})}
\newcommand{\Eqref}[1]{Eq.\ (\ref{#1})}

\newcommand{\Eqsandref}[2]{Eqs.\ (\ref{#1}) and (\ref{#2})}
\newcommand{\figref}[1]{Fig.\ \ref{#1}}

\newcommand{\bea}{\begin{eqnarray}}
\newcommand{\eea}{\end{eqnarray}}
\newcommand{\lt}{\left}
\newcommand{\rt}{\right}
\newcommand{\bl}{\bigl}
\newcommand{\br}{\bigr}
\newcommand{\la}{\langle}
\newcommand{\ra}{\rangle}

\newcommand{\dd}{\partial}
\newcommand{\vdel}{\boldsymbol{\nabla}}
\newcommand{\dpar}{{\nabla}_{\parallel}}
\newcommand{\delsq}{\mathrm{\Delta}}

\newcommand{\vx}{\mathbf{x}}

\newcommand{\vk}{\mathbf{k}}
\newcommand{\tk}{\tilde k}
\newcommand{\tvk}{\tilde\vk}
\newcommand{\vu}{\mathbf{u}}
\newcommand{\vo}{\boldsymbol{\omega}}
\newcommand{\vW}{\boldsymbol{\Sigma}}
\newcommand{\vf}{\mathbf{f}}
\newcommand{\vB}{\mathbf{B}}
\newcommand{\vK}{\mathbf{K}}
\newcommand{\vb}{\hat{\mathbf{b}}}
\newcommand{\bb}{\hat{b}}

\newcommand{\unity}{\hat{\mathsf{I}}}
\newcommand{\vz}{\mathbf{z}}
\newcommand{\tB}{\tilde B}
\newcommand{\tvB}{\tilde\vB}
\newcommand{\vM}{\hat{\mathsf{M}}}
\newcommand{\vR}{\hat{\mathsf{R}}}
\newcommand{\vL}{\hat{\mathsf{L}}}
\newcommand{\ve}{\hat{\mathbf{e}}}

\newcommand{\du}{\delta u}
\newcommand{\dB}{\delta B}
\newcommand{\dz}{\delta z}
\newcommand{\dvB}{\delta\vB}
\newcommand{\dvu}{\delta\vu}
\newcommand{\dvuperp}{\delta\vu_{\perp}}
\newcommand{\drho}{\delta\rho}
\newcommand{\dpr}{\delta p}
\newcommand{\ds}{\delta s}
\newcommand{\Es}{{I}_{s}}
\newcommand{\Eperp}{{I}_{\perp}}
\newcommand{\Epar}{{I}_{\parallel}}

\newcommand{\vuperp}{\vu_{\perp}}
\newcommand{\uperp}{{u}_{\perp}}
\newcommand{\upar}{{u}_{\parallel}}
\newcommand{\zpar}{{z}_{\parallel}}
\newcommand{\dvBperp}{\dvB_{\perp}}
\newcommand{\dBperp}{\delta B_{\perp}}
\newcommand{\dBpar}{\delta B_{\parallel}}
\newcommand{\Dpar}{{\vB\over B_{0}}\cdot\vdel}
\newcommand{\vdperp}{\vdel_{\perp}}
\newcommand{\dperp}{{\nabla}_{\perp}}

\newcommand{\diag}{\mathrm{diag}}
\newcommand{\const}{\mathrm{const}}

\newcommand{\uA}{{v}_{\mathrm{A}}}
\newcommand{\tA}{{\tau}_{\mathrm{A}}}
\newcommand{\teddy}{{\tau}_{\mathrm{s}}}
\newcommand{\kpar}{{k}_{\parallel}}
\newcommand{\kperp}{{k}_{\perp}}
\newcommand{\Lpar}{{L}_{\parallel}}
\newcommand{\kparo}{{k}_{\parallel0}}
\newcommand{\lpar}{{l}_{\parallel}}
\newcommand{\lperp}{{l}_{\perp}}
\newcommand{\ld}{{l}_{\nu}}
\newcommand{\ls}{{l}_{\mathrm{s}}}
\newcommand{\lres}{{l}_{\eta}}
\newcommand{\mfp}{{\lambda}_{\mathrm{mfp}}}
\newcommand{\Rey}{{Re}}
\newcommand{\Rm}{{R}_{m}}
\newcommand{\Rmc}{{R}_{m,c}}
\newcommand{\Pm}{{Pr}_{m}}

\newcommand{\Bsq}{\la B^2\ra}
\newcommand{\usq}{\la u^2\ra}
\newcommand{\vth}{{v}_{\mathrm{th},i}}
\newcommand{\vperp}{{v}_{\perp}}

\newcommand{\nui}{{\nu}_{ii}}
\newcommand{\nupar}{{\nu}_{\parallel}}
\newcommand{\pperp}{{p}_{\perp}}
\newcommand{\ppar}{{p}_{\parallel}}

\begin{document}

\title*{Turbulence and Magnetic Fields in Astrophysical Plasmas}
\author{Alexander A Schekochihin\inst{1} and Steven C Cowley\inst{2}}
\institute{
DAMTP, University of Cambridge, Cambridge CB3 0WA 
and Department of Physics, Imperial College, London SW7 2BW, United Kingdom
(\texttt{a.schekochihin@imperial.ac.uk})
\and
Department of Physics and Astronomy, UCLA, Los Angeles, CA 90095-1547, USA 
and Department of Physics, Imperial College, London SW7 2BW, United Kingdom
(\texttt{cowley@physics.ucla.edu})  
}
\maketitle

\section{Introduction}
Magnetic fields permeate the Universe. 
They are found in planets, stars, accretion discs, galaxies, 
clusters of galaxies, and the intergalactic medium. 
While there is often a component of the field that is 
spatially coherent at the scale of the astrophysical 
object, the field lines 
are tangled chaotically and there are magnetic fluctuations 
at scales that range over orders of magnitude. 
The cause of this disorder is the turbulent state of 
the plasma in these systems. 
This plasma is, as a rule, highly conducting, 
so the magnetic field lines are entrained by (frozen into) the 
fluid motion. As the fields are stretched and bent 
by the turbulence, they can resist deformation by exerting 
the Lorentz force on the plasma. 
The turbulent advection of the magnetic field and the field's back reaction 
together give rise to the statistically steady state of 
fully developed MHD turbulence. 
In this state, energy and momentum injected at large 
(object-size) scales are transfered to smaller scales and 
eventually dissipated. 

Despite over fifty years of research and many major advances,
a satisfactory theory of MHD turbulence remains elusive.  Indeed, 
even the simplest (most idealised) cases are still not fully understood. 
One would hope that there are universal properties of MHD turbulence that 
hold in all applications --- or at least in a class of applications.  
Among the most important questions for astrophysics that a successful theory 
of turbulence must answer are:
\begin{itemize}
\item How does the turbulence amplify, sustain and shape magnetic fields? 
What is the structure and spectrum of this field at large and small scales? 
The problem of turbulence in astrophysics is thus directly related to the 
fundamental problem of magnetogenesis. 
\item How is energy cascaded and dissipated in plasma turbulence?  
In accretion discs and the solar corona, for example, 
one would like to know if the turbulence heats ions or electrons 
predominantly \cite{Quataert_Gruzinov_heating}. 
\item How does the turbulent flow and magnetic field enhance or inhibit the 
transport of heat, (angular) momentum, and cosmic rays? Again in accretion 
discs, a key parameter is the effective turbulent viscosity that causes 
the transport of angular momentum and enables accretion \cite{Shakura_Sunyaev}. 
In cluster physics, an understanding of how viscous heating and thermal conduction 
in a turbulent magnetised plasma balance the radiative cooling 
is necessary to explain the observed global temperature profiles 
\cite{Dennis_Chandran}. 
\end{itemize}

%Thus, the theory of MHD 
%turbulence describes the transport properties of the 
%turbulent matter that makes up the astrophysical objects 
%and is, therefore, essential for understanding their 
%physics both at small {\em and} large scales. 
%Furthermore, it is the turbulence that has almost certainly amplified, 
%sustained and shaped the field that we see. The problem of 
%astrophysical turbulence is thus directly related to the 
%fundamental problem of magnetogenesis. 

In this chapter, we discuss the current understanding of the most basic 
properties of astrophysical MHD turbulence. We emphasise possible universal 
aspects of the theory. We shall touch primarily 
on two applications: turbulence in the solar wind and in clusters of 
galaxies. These are, in a certain (very approximate) sense, 
two ``pure'' cases of small-scale turbulence, where theoretical models 
of the two main regimes of MHD turbulence (discussed in \secref{sec_aniso} 
and \secref{sec_iso}) can be put to the test. They are also 
good examples of a complication that is more or less generic 
in astrophysical plasmas: the MHD description is, in fact, insufficient 
for astrophysical turbulence and plasma physics must make an entrance. 
Why this is so will be explained in \secref{sec_plasma}. 

The astrophysical plasma turbulence is even more of a {\it terra incognita} 
than the MHD turbulence, so we shall start with the equations of 
incompressible MHD --- the simplest equations that describe 
(subsonic) turbulent dynamics in a conducting medium:
\bea
\label{NSEq}
{d\vu\over dt} &=& -\vdel p + \nu\delsq\vu + 
\vB\cdot\vdel\vB + \vf,\quad\vdel\cdot\vu=0,\\
{d\vB\over dt} &=& \vB\cdot\vdel\vu + \eta\delsq\vB,
\label{ind_eq}
\eea
where $\vu$ is the velocity field, 
$d/dt=\dd_t+\vu\cdot\vdel$ the convective derivative, 
$p$ the pressure (scaled by the constant density $\rho$ and 
determined by the incompressibility constraint), 
$\vB$ the magnetic field scaled by $(4\pi\rho)^{1/2}$, 
$\nu$ the kinematic viscosity, $\eta$ the magnetic diffusivity, 
and $\vf$ the body force that models large-scale energy input. 
The specific energy injection mechanisms vary: typically, in astrophysics, 
these are either background gradients (e.g., the temperature gradient in 
stellar convective zones, the Keplerian velocity shear in accretion 
discs), which mediate the conversion of gravitational energy 
into kinetic energy of fluid motion, or direct sources of energy 
such as the supernovae in the interstellar medium or active 
galactic nuclei in galaxy clusters. 
What all these injection mechanisms have in common is 
that the scale at which they operate, hereafter denoted by $L$, 
is large, comparable with the size of the system. 
While the large-scale dynamics depend on the specific 
astrophysical situation, it is common to assume that, once 
the energy has cascaded down to scales substantially smaller than $L$, 
the nonlinear dynamics are universal. 

The universality of small scales is a cornerstone of all theories 
of turbulence. It goes back to Kolmogorov's 1941 dimensional 
theory (or K41 \cite{K41}; 
see \cite{Landau_Lifshitz}, $\S33$ for a lucid and concise exposition). 
Here is an outline Kolmogorov's reasoning. 
Consider \eeqref{NSEq} without the magnetic term. 
Denote the typical fluctuating velocity difference across 
scale $L$ by $\du_L$. The energy associated with these fluctuations 
is $\du_L^2$ and the characteristic time for this energy to cascade 
to smaller scales by nonlinear coupling is $L/\du_L$. The total specific 
power (energy flux) going into the turbulent cascade is then 
$\epsilon=\la\vu\cdot\vf\ra\sim\du_L^3/L$. 
In a statistically stationary situation, all this power must be dissipated,
so $\epsilon=\nu\la|\vdel\vu|^2\ra$. Since $\epsilon$ is a finite 
quantity completely defined by the large-scale energy-injection 
process, it cannot depend on $\nu$. For very small $\nu$, this 
implies that the velocity must develop very small scales 
so that $\nu\la|\vdel\vu|^2\ra$ has a constant limit as $\nu\to+0$. 
The only quantity with dimensions of length that one 
can construct out of $\epsilon$ and $\nu$ is 
$\ld\sim(\nu^3/\epsilon)^{1/4}\sim \Rey^{-3/4}L$, where 
$\Rey\sim\du_L L/\nu$ is the Reynolds number. 
In astrophysical applications, $\Rey$ is usually large, 
so the viscous dissipation occurs at scales $\ld\ll L$. 
The energy injected at the large scale $L$ must be transfered 
to the small scale $\ld$ across a range of scales (the inertial range). 
The hydrodynamic turbulence theory assumes that the physics 
in this range is universal, i.e., it
depends neither on the energy-injection mechanism nor 
on the dissipation mechanism. 
Four further assumptions are made about the inertial range: 
{\bf homogeneity} (no special points), 
{\bf scale invariance} (no special scales), 
{\bf isotropy} (no special directions), 
and {\bf locality of interactions} 
(interactions between comparable scales dominate). 
Then, at each scale $l$ such that $L\gg l\gg \ld$, 
the total power $\epsilon$ must arrive from larger scales 
and be passed on to smaller scales:
\bea
\label{eps_general}
\epsilon \sim \du_l^2/\tau_l,
\eea 
where $\du_l$ is the velocity difference across scale $l$ and 
$\tau_l$ the cascade time. Dimensionally, only one time scale 
can be constructed out of the local quantities $\du_l$ and $l$: 
$\tau_l\sim l/\du_l$. Substituting this into \eeqref{eps_general} 
and solving for $\du_l$, we arrive at Kolmogorov's scaling: 
$\du_l\sim(\epsilon l)^{1/3}$, or, for the energy spectrum $E(k)$, 
\bea
\du_l^2\sim\int_{k=1/l}^\infty dk'E(k')\sim\epsilon^{2/3} k^{-2/3}
\quad\Rightarrow\quad
E(k)\sim\epsilon^{2/3} k^{-5/3}. 
\label{E_K41}
\eea

The history of the theory of MHD turbulence over the last half century has 
been that of a succession of attempts to adapt the K41-style thinking 
to fluids carrying magnetic fields. In the next two sections, 
we give an overview of these efforts and of the resulting gradual 
realisation that the key assumptions of the small-scale universality, isotropy 
and locality of interactions fail in various MHD contexts. 

\section{Alfv\'enic turbulence} 
\label{sec_aniso}
Let us consider the case of a plasma threaded by 
a straight uniform magnetic field $\vB_0$ of some external 
(i.e., large-scale) origin. Let us also consider weak forcing so that 
the fundamental turbulent excitations are small-amplitude wave-like 
disturbances propagating along the mean field.  We will refer to such a 
limit as Alfv\'enic turbulence --- it is manifestly anisotropic.

\subsection{Iroshnikov--Kraichnan turbulence}
\label{sec_IK}

If we split the magnetic field 
into the mean and fluctuating parts, $\vB=\vB_0 + \dvB$, 
and introduce Elsasser \cite{Elsasser} variables $\vz^\pm=\vu\pm\dvB$, 
\eqsandref{NSEq}{ind_eq} take a symmetric form: 
\bea
\dd_t\vz^\pm \mp \uA\dpar\vz^\pm + \vz^\mp\cdot\vdel\vz^\pm 
= -\vdel p + {\nu+\eta\over 2}\delsq\vz^\pm + 
{\nu-\eta\over 2}\delsq\vz^\mp + \vf, 
\label{eq_z}
\eea
where $\uA=|\vB_0|$ is the Alfv\'en speed 
and $\dpar$ is the gradient in the direction of the 
mean field $\vB_0$. The Elsasser equations have a simple {\em exact} solution: 
if $\vz^+=0$ or $\vz^-=0$, the nonlinear term vanishes 
and the other, non-zero, Elsasser field is simply 
a fluctuation of arbitrary shape and magnitude 
propagating along the mean field at the Alfv\'en speed $\uA$.
Kraichnan \cite{Kraichnan} realised in 1965 that the Elsasser form 
of the MHD equations only allows nonlinear interactions between 
counterpropagating such fluctuations. 
The phenomenological theory that he and, independently, Iroshnikov \cite{Iroshnikov}, 
developed on the basis of this idea (the IK theory) can be summarised as follows. 

Following the general philosophy of K41, assume that 
only fluctuations of comparable scales interact ({\bf locality of interactions})
and consider these interactions in 
the inertial range, comprising scales $l$ smaller than the forcing scale 
$L$ and larger than the (still to be determined) dissipation scale. 
Let us think of the fluctuations propagating in either direction 
as trains of spatially localised Alfv\'en-wave\footnote{Waves in incompressible 
MHD can have either the Alfv\'en- or the slow-wave polarisation. Since both 
propagate at the Alfv\'en speed, we shall, for simplicity, refer to them 
as Alfv\'en waves. The differences between the Alfv\'en- and slow-wave 
cascades are explained in detail at the end of \secref{sec_GS}.} packets of 
parallel (to the mean field) extent $\lpar$ and perpendicular extent $l$ 
(we shall not, for the time being, specify how $\lpar$ relates to $l$). 
Assume further that $\dz^+_l\sim\dz^-_l\sim\du_l\sim\dB_l$. 
We can again use \eeqref{eps_general} for the energy flux through 
scale $l$, but there is, unlike in the case of purely hydrodynamic 
turbulence, no longer 
a dimensional inevitability about the determination of the 
cascade time $\tau_l$ because two physical time scales are associated 
with each wave packet: the Alfv\'en time $\tA(l)\sim\lpar/\uA$ 
and the strain (or ``eddy'') time $\teddy(l)\sim l/\du_l$. 
To state this complication in a somewhat more formal way, 
there are three dimensionless combinations 
in the problem of MHD turbulence: 
$\epsilon l/\du_l^3$, 
$\du_l/\uA$ and $\lpar/l$, so 
the dimensional analysis does not uniquely determine scalings 
and further physics input is needed. 

Two counterpropagating wave packets take an Alfv\'en 
time to pass through each other. 
During this time, the amplitude of either packet 
is changed by 
\bea
\Delta\du_l\sim {\du_l^2\over l}\tA\sim\du_l{\tA\over \teddy}.
\eea
The IK theory now assumes {\bf weak interactions}, 
$\Delta\du_l\ll\du_l\Leftrightarrow\tA\ll\teddy$. 
The cascade time $\tau_l$ is estimated as the time it takes 
(after many interactions) to change $\du_l$ by an amount comparable 
to itself. If the changes in amplitude accumulate like a random walk, 
we have
\bea
\label{t_cascade}
\sum^t\Delta\du_l \sim \du_l{\tA\over \teddy}\sqrt{t\over \tA} \sim \du_l
\quad\mathrm{for}\quad t\sim\tau_l
\quad\Rightarrow\quad
\tau_l\sim {\teddy^2\over \tA}\sim {l^2\uA\over \lpar\du_l^2}.
\eea
Substituting the latter formula into \eeqref{eps_general}, 
we get 
\bea
\label{du_general}
\du_l\sim (\epsilon\uA)^{1/4}\lpar^{-1/4} l^{1/2}.
\eea
The final IK assumption, which at the time seemed reasonable in light 
of the success of the K41 theory, was that of {\bf isotropy}, 
fixing the dimensionless ratio $\lpar/l\sim1$, and, therefore, the scaling:
\bea
\du_l\sim (\epsilon\uA)^{1/4} l^{1/4}
\quad\Rightarrow\quad
E(k)\sim (\epsilon\uA)^{1/2} k^{-3/2}. 
\eea

\subsection{Turbulence in the solar wind}
\label{sec_swind}

The solar wind, famously predicted by Parker \cite{Parker_solar_wind}, 
was the first astrophysical plasma in which direct 
measurements of turbulence became possible \cite{Coleman}. 
A host of subsequent observations (for a concise review, see 
\cite{Goldstein_Roberts}) revealed powerlike 
spectra of velocity and magnetic fluctuations 
in what is believed to be the inertial 
range of scales extending roughly from $10^6$ to $10^3$ km. 
The mean magnetic field is $B_0\sim10...10^2\ \mu$G, while 
the fluctuating part $\dB$ is a factor of a few smaller. 
The velocity dispersion is $\du\sim10^2$ km/s, approximately 
in energy equipartition with $\dB$. 
The $\vu$ and $\vB$ fluctuations are highly correlated 
at all scales and almost undoubtedly Alfv\'enic. 
It is, therefore, natural to think 
of the solar wind as a space laboratory conveniently at our 
disposal to test theories of Alfv\'enic turbulence in astrophysical conditions. 

For nearly 30 years following Kraichnan's paper \cite{Kraichnan}, 
the IK theory was accepted as the correct extension of K41 to MHD turbulence 
and, therefore, with minor modifications allowing for the observed imbalance between 
the energies of the $\vz^+$ and $\vz^-$ fluctuations \cite{Dobrowolny_Mangeney_Veltri}, 
also to the turbulence in the solar wind. 
However, alarm bells were sounding already in 1970s and 1980s when 
measurements of the solar-wind turbulence 
revealed that it was strongly anisotropic with $\lpar>\lperp$ 
\cite{Belcher_Davis} (see also \cite{Matthaeus_Goldstein_Roberts}) 
and that its spectral index was closer to $-5/3$ than to $-3/2$ 
\cite{Matthaeus_Goldstein}.\footnote{\label{fn_ARS}
Another classic example of a $-5/3$ scaling 
in astrophysical turbulence is the spectrum of electron 
density fluctuations (thought to trace the velocity spectrum) 
in the interstellar medium --- the famous ``power law in the sky,'' 
which appears to hold across 12 decades of scales 
\cite{Armstrong_Rickett_Spangler}.}
Numerical simulations have 
confirmed the anisotropy of MHD turbulence in the presence of a strong 
mean field \cite{Maron_Goldreich,Cho_Lazarian_Vishniac,Mueller_Biskamp_Grappin}. 

\subsection{Weak turbulence}
\label{sec_weak}

The realisation that the isotropy assumption must be abandoned 
led to a reexamination of the Alfv\'en-wave interactions in MHD 
turbulence. If the assumption of weak interactions is kept, 
MHD turbulence can be regarded as an ensemble of waves, whose 
wavevectors $\vk$ and frequencies $\omega^\pm(\vk)=\pm\kpar\uA$ have 
to satisfy resonance conditions in order for an interaction to occur. 
For three-wave interactions (1 and 2 counterpropagating, 
giving rise to 3), 
\bea
\vk_1&+&\vk_2=\vk_3 \qquad\qquad\,\quad\Rightarrow\quad  
k_{\parallel1}+k_{\parallel2}=k_{\parallel3},\\
\omega^\pm(\vk_1)&+&\omega^\mp(\vk_2)=\omega^\pm(\vk_3) \quad\Rightarrow\quad
k_{\parallel1} - k_{\parallel2} = k_{\parallel3},
\eea
whence $k_{\parallel2}=0$ and $k_{\parallel3}=k_{\parallel1}$. 
Thus (i) interactions do not change $\kpar$; 
(ii) interactions are mediated by modes with $\kpar=0$, 
which are quasi-2D fluctuations rather than waves 
\cite{Montgomery_Turner,Shebalin_Matthaeus_Montgomery,Ng_Bhattacharjee96}.\footnote{Goldreich \& Sridhar 
\cite{GS97} argued that the time it takes three waves to realise that one 
of them has zero frequency is infinite and, therefore, the weak-interaction
approximation cannot be used. This difficulty can, in fact, be removed 
by noticing that the $\kpar=0$ modes have a finite correlation 
time, but we do not have space to discuss this rather subtle issue here 
(two relevant references are \cite{Galtier_etal,Lithwick_Goldreich03}).} 

The first of these conclusions suggests a quick fix 
of the IK theory: take $\lpar\sim\kparo^{-1}=\const$ 
(the wavenumber at which the waves are launched) and $l\sim\lperp$ in 
\eeqref{du_general} ({\bf no parallel cascade}). Then 
the spectrum is \cite{GS97}
\bea
E(\kperp) \sim (\epsilon\kparo\uA)^{1/2}\kperp^{-2}.
\label{E_WT}
\eea
The same result can be obtained via a formal 
calculation based on the standard weak-turbulence 
theory \cite{Galtier_etal,Lithwick_Goldreich03}. 
However, it is not uniformly valid at all $\kperp$. 
Indeed, let us check if the assumption of weak interactions, $\tA\ll\teddy$, 
is actually satisfied by the scaling relation \exref{du_general} 
with $\lpar\sim\kparo^{-1}$:
\bea
\label{WT_range}
{\tA\over \teddy}\sim{\epsilon^{1/4}\over \bl(\kparo\uA\br)^{3/4}\lperp^{1/2}}\ll1
\quad\Leftrightarrow\quad
\lperp\gg l_*={\epsilon^{1/2}\over \bl(\kparo\uA\br)^{3/2}}
\sim{\du_L^2\over \uA^2}{1\over \kparo^2L}, 
\eea 
where $\du_L$ is the velocity at the outer scale (the rms velocity). 
Thus, if $\Rey=\du_L L/\nu$ and $\Rm=\du_L L/\eta$ are large enough, 
the inertial range will always contain a scale $l_*$ below which 
the interactions are no longer weak.\footnote{There is also an upper limit 
to the scales at which \eeqref{E_WT} is applicable. 
The boundary conditions at the ends of the ``box'' are unimportant only 
if the cascade time \exref{t_cascade} is shorter than the time it takes 
an Alfv\'enic fluctuation to cross the box:
$\tau_l\ll\Lpar/\uA\ \Leftrightarrow\  
\lperp\ll L_*=(\epsilon/\kparo\uA^3)^{1/2}\Lpar$, where
$\Lpar$ is the length of the box along the mean field.
Demanding that $L_*>L$, the perpendicular size of the box, 
we get a lower limit on the aspect ratio of the box: 
$\Lpar/L > \kparo L(\uA/\du_L)^2$. If this is not satisfied, 
the physical (non-periodic) boundary conditions may impose 
a limit on the perpendicular field-line wander and thus effectively 
forbid the $\kpar=0$ modes. It has been suggested \cite{GS97} that 
a weak-turbulence theory based on 4-wave interactions \cite{Sridhar_Goldreich} 
should then be used at $\lperp>L_*$. \label{fn_zero_modes}} 

\subsection{Goldreich--Sridhar turbulence}
\label{sec_GS}

In 1995, Goldreich \& Sridhar \cite{GS95} conjectured that the 
strong turbulence below the scale $l_*$ should satisfy 
\bea
\tA\sim\teddy\quad\Leftrightarrow\quad
\lpar/\lperp\sim\uA/\du_l, 
\label{crit_bal}
\eea
a property that has come to be known as {\bf the critical balance}. 
Goldreich \& Sridhar argued that when $\tA\ll\teddy$, the weak turbulence 
theory "pushes" the spectrum towards the approximate equality \exref{crit_bal}. 
They also argued that when $\tA\gg \teddy$, motions along 
the field lines are decorrelated and naturally 
develop the critical balance. 

\begin{figure}[t]
\centering
\begin{tabular}{ccc}
\includegraphics[height=2.5cm]{iroshnikov.epsf} & &
\includegraphics[height=2.5cm]{kraichnan.epsf}\\ 
%http://facultyprofiles.engineering.jhu.edu/ME/listing.html?select=fl&id=260&item=g
R.\ S.\ Iroshnikov (1937-1991) & & R.\ H.\ Kraichnan\\
& {\qquad} &\\
\includegraphics[height=2.35cm]{goldreich.epsf} & & 
%http://www.gps.caltech.edu/faculty/goldreich/
\includegraphics[height=2.35cm]{sridhar.epsf}\\
P.\ Goldreich & & S.\ Sridhar\\
\end{tabular}
\caption{IK and GS (photo of R.\ S.\ Iroshnikov courtesy 
of Sternberg Astronomical Institute; photo of R.\ H.\ Kraichnan 
courtesy of the Johns Hopkins University).} 
\label{fig_IKGS}
\end{figure}

The critical balance fixes the relation between 
two of the three dimensionless combinations in MHD turbulence. 
Since now there is only one natural time scale associated with fluctuations 
at scale $l$, this time scale is now 
assumed to be the cascade time, $\tau_l\sim\teddy$. 
This brings back Kolmogorov's spectrum \exref{E_K41} for 
the perpendicular cascade. The parallel cascade is now also 
present but is weaker: from \eeqref{crit_bal}, 
\bea
\lpar\sim\uA\epsilon^{-1/3}\lperp^{2/3}\sim\kparo^{-1}\bl(\lperp/l_*\br)^{2/3}.
\label{GS_cone}
\eea

The scalings \exsandref{E_K41}{GS_cone} should hold at all 
scales $\lperp\ll l_*$ and above the dissipation 
scale: either viscous $\ld\sim(\nu^3/\epsilon)^{1/4}$ 
or resistive $\lres\sim(\eta^3/\epsilon)^{1/4}$, 
whichever is larger. Comparing these scales with $l_*$ [\eeqref{WT_range}], 
we note that the strong-turbulence range is non-empty only if 
$\Rey, \Rm \gg \bl(\kparo L\br)^3\bl(\uA/\du_L\br)^3$,
a condition that is effortlessly satisfied in most astrophysical 
cases but should be kept in mind when numerical simulations are 
undertaken. 

The Goldreich--Sridhar (GS) theory has now replaced the IK theory 
as the standard accepted description of MHD turbulence. 
The feeling that the GS theory is the right one, 
created by the solar wind \cite{Matthaeus_Goldstein} and 
ISM \cite{Armstrong_Rickett_Spangler} observations 
that show a $k^{-5/3}$ spectrum, 
is, however, somewhat spoiled by the consistent failure of the numerical 
simulations to produce such a spectrum
\cite{Maron_Goldreich,Mueller_Biskamp_Grappin}. Instead, a spectral 
index closer to IK's $-3/2$ is obtained (this seems to be 
the more pronounced the stronger the mean field), 
although the turbulence is definitely anisotropic and the GS 
relation \exref{GS_cone} appears to be satisfied 
\cite{Maron_Goldreich,Cho_Lazarian_Vishniac}! 
This trouble has been blamed on intermittency 
\cite{Maron_Goldreich}, a perennial scapegoat of turbulence theory, 
but a non-speculative solution remains to be found. 

The puzzling refusal of the numerical MHD turbulence to agree 
with either the GS theory or, indeed, with the solar-wind observations 
highlights the rather shaky quality of the existing 
physical understanding of what really happens in a turbulent 
magnetic fluid on the dynamical level. 
One conceptual difference between MHD and hydrodynamic turbulence is the possibility 
of long-time correlations. In the large-$\Rm$ limit, the magnetic field is determined 
by the displacement of the plasma, i.e., the time integral of the (Lagrangian) 
velocity. In a stable plasma, the field-line tension tries to return the field line 
to the unperturbed equilibrium position. Only ``interchange'' ($\kpar=0$) motions of the 
{\em entire} field lines are not subject to this ``spring-back'' effect. 
Such motions are often ruled out by geometry or boundary conditions 
(cf.\ footnote \ref{fn_zero_modes}). Thus, fluid elements 
in MHD cannot simply random walk as this would increase (without bound) the field-line 
tension. However, they may random walk for a substantial period before 
the tension returns them back to the equilibrium state. The role of such 
long-time correlations in MHD turbulence is unknown.\\ 

{\small

\noindent
{\bf Reduced MHD, the decoupling of the Alfv\'en-wave cascade, and 
turbulence in the interstellar medium.} We now give a rigourous 
demonstration of how the turbulent cascade associated with the Alfv\'en waves 
(or, more precisely, Alfv\'en-wave-polarised fluctuations) decouples from the 
cascades of the slow waves and 
entropy fluctuations. Let us start with the equations 
of compressible MHD:
\bea
\label{MHD_rho}
{d\rho\over dt} &=& -\rho\vdel\cdot\vu,\\
\label{MHD_u}
\rho\,{d\vu\over dt} &=& -\vdel\lt(p+{B^2\over 8\pi}\rt) + {\vB\cdot\vdel\vB\over 4\pi},\\
\label{MHD_p}
{ds\over dt} &=& 0,\quad s={p\over \rho^\gamma},\quad \gamma={5\over 3},\\
\label{MHD_B}
{d\vB\over dt} &=& \vB\cdot\vdel\vu - \vB\vdel\cdot\vu.
\eea
Consider a uniform static equilibrium with a straight magnetic field, 
so $\rho = \rho_0 + \drho$, 
$p = p_0 + \dpr$,
$\vB = \vB_0 + \dvB$. 
Based on observational and numerical evidence, it is safe 
to assume that the turbulence in such a system will be anisotropic 
with $\kpar\ll\kperp$. Let us, therefore, introduce a small parameter 
$\epsilon\sim\kpar/\kperp$ and carry out a systematic expansion 
of \eqsref{MHD_rho}{MHD_B} in $\epsilon$. In this expansion, the 
fluctuations are treated as small, but not arbitrarily so: 
in order to estimate their size, 
we shall adopt the critical-balance conjecture \exref{crit_bal}, 
which is now treated not as a detailed scaling prescription but as an 
ordering assumption. This allows us to introduce the following ordering:
\bea
{\drho\over \rho_0}
\sim {\uperp\over v_A} \sim {\upar\over v_A}
\sim {\dpr\over p_0}
\sim {\dBperp\over B_0} \sim {\dBpar\over B_0} 
\sim {\kpar\over \kperp} 
\sim \epsilon, 
\label{RMHD_ordering}
\eea
where we have also assumed that the velocity and magnetic-field
fluctuations have the character of Alfv\'en and slow waves 
($\dvB\sim\vu$) and that the relative amplitudes of the 
Alfv\'en-wave-polarised fluctuations ($\uperp/v_A$, $\dBperp/B_0$), 
slow-wave-polarised fluctuations ($\upar/v_A$, $\dBpar/B_0$) 
and density fluctuations ($\drho/\rho_0$) are all the same 
order.\footnote{Strictly speaking, whether this is the case 
depends on the energy sources that drive the turbulence: as we are  
about to see, if no slow waves are launched, none will be present. 
However, it is safe to assume in astrophysical contexts that the 
large-scale energy input is random and, therefore, comparable power 
is injected in all types of fluctuations.} 
We further assume that the characteristic frequency of the 
fluctuations is $\omega\sim\kpar v_A$, which means that 
the fast waves, for which $\omega\simeq \kperp\sqrt{v_A^2+c_s^2}$, 
where $c_s=\gamma p_0/\rho_0$ 
is the sound speed, are ordered out. 

We start by observing that 
the Alfv\'en-wave-polarised fluctuations are two-dimensionally solenoidal: since 
$\vdel\cdot\vu=O(\epsilon^2)$ [from \eeqref{MHD_rho}] and $\vdel\cdot\vB=0$, 
separating the $O(\epsilon)$ part of these divergences gives 
$\vdperp\cdot\vuperp=\vdperp\cdot\dvBperp=0$. 
We may, therefore, express $\vuperp$ and $\dvBperp$ in terms of 
scalar stream (flux) functions:
\bea
\vuperp = \vb_0\times\vdperp\phi,\qquad 
{\dvBperp\over \sqrt{4\pi\rho_0}} = \vb_0\times\vdperp\psi,
\label{phi_psi_def}
\eea
where $\vb_0=\vB_0/B_0$. 
Evolution equations for $\phi$ and $\psi$ 
are obtained by substituting the expressions \exref{phi_psi_def} into 
the perpendicular parts of the induction equation \exref{MHD_B} and the 
momentum equation \exref{MHD_u} --- of the latter the curl is taken to 
annihilate the pressure term. Keeping only the terms of the lowest order, 
$O(\epsilon^2)$, we get 
\bea
\label{RMHD_psi}
{\dd\over \dd t}\,\psi + \lt\{\phi,\psi\rt\} &=& v_A\dpar\phi,\\
\label{RMHD_phi}
{\dd\over \dd t}\dperp^2\phi + \lt\{\phi,\dperp^2\phi\rt\} 
&=& v_A\dpar\dperp^2\psi + \lt\{\psi,\dperp^2\psi\rt\},
\eea
where $\lt\{\phi,\psi\rt\}=\vb_0\cdot(\vdperp\phi\times\vdperp\psi)$ 
and, to lowest order,  
\bea
\label{dt_def}
{d\over dt} &=& {\dd\over \dd t} + \vuperp\cdot\vdperp={\dd\over \dd t} + \lt\{\phi,\cdots\rt\},\\ 
\label{dpar_def}
\Dpar &=& \dpar + {\dvBperp\over B_0}\cdot\vdperp 
= \dpar + {1\over v_A}\lt\{\psi,\cdots\rt\}.
\eea
\Eqsandref{RMHD_psi}{RMHD_phi} are known as the Reduced Magnetohydrodynamics 
(RMHD). They were first derived by Strauss \cite{Strauss} 
in the context of fusion plasmas. 
They form a closed set, meaning that the Alfv\'en-wave cascade 
decouples from the slow waves and density fluctuations. 

In order to derive evolution equations for the latter, 
let us revisit the perpendicular part of the momentum equation and 
use \eeqref{RMHD_ordering} to order terms in it.
In the lowest order, $O(\epsilon)$, we get the pressure balance
\bea
\label{MHD_pr_bal}
\vdperp\lt(\dpr + {B_0\dBpar\over 4\pi}\rt) = 0 \quad\Rightarrow\quad
{\dpr\over p_0} = -\gamma\,{v_A^2\over c_s^2}{\dBpar\over B_0}.
\eea
Using \eeqref{MHD_pr_bal} and the entropy 
equation \exref{MHD_p}, we get 
\bea
{d\over dt}{\ds\over s_0} = 0,\quad 
{\ds\over s_0} = {\dpr\over p_0} - \gamma{\drho\over \rho_0} = 
- \gamma\lt({\drho\over \rho_0} + {v_A^2\over c_s^2}{\dBpar\over B_0}\rt),
\label{eq_ds}
\eea
where $s_0=p_0/\rho_0^\gamma$. 
On the other hand, from the continuity equation \exref{MHD_rho} and 
the parallel component of the induction equation \exref{MHD_B}, 
\bea
\label{eq1}
{d\over dt}\lt({\drho\over \rho_0} - {\dBpar\over B_0}\rt) + \Dpar\upar = 0.
\eea
Combining \eqsandref{eq_ds}{eq1}, we obtain 
\bea
\label{eq_drho}
{d\over dt}{\drho\over \rho_0} &=& - {1\over 1+ c_s^2/v_A^2}\,\Dpar\upar,\\
\label{eq_Bpar}
{d\dBpar\over dt} &=& {1\over 1 + v_A^2/c_s^2}\,\Dpar\upar.
\eea
Finally, we take the parallel component of the momentum 
equation \exref{MHD_u} and notice that, due to \eeqref{MHD_pr_bal} and to 
the smallness of the parallel gradients, the pressure term is $O(\epsilon^3)$, 
while the inertial and tension terms are $O(\epsilon^2)$. Therefore, 
\bea
\label{eq_upar}
{d\upar\over dt} = v_A^2\Dpar{\dBpar\over B_0}.
\eea 
\Eqsandref{eq_Bpar}{eq_upar} describe the slow-wave-polarised fluctuations, 
while \eeqref{eq_ds} describes the zero-frequency entropy mode. 
The nonlinearity in these equations enters 
via the derivatives defined in \eqsref{dt_def}{dpar_def} and is due solely to 
interactions with Alfv\'en waves. 

Naturally, the reduced equations derived above can be cast in the Elsasser form. 
If we introduce Elsasser potentials $\zeta^\pm=\phi\pm\psi$, \eqsandref{RMHD_psi}{RMHD_phi} 
become 
\bea
{\dd\over \dd t}\dperp^2\zeta^\pm \mp 
v_A\dpar\dperp^2\zeta^\pm 
= -{1\over 2}\lt[\lt\{\zeta^+,\dperp^2\zeta^-\rt\}
+ \lt\{\zeta^-,\dperp^2\zeta^+\rt\}
\mp\dperp^2\lt\{\zeta^+,\zeta^-\rt\}\rt].\quad
\label{eq_zeta}
\eea
This is the same as the perpendicular part of \eeqref{eq_z} 
with $\vz^\pm_\perp=\vb_0\times\vdperp\zeta^\pm$. 
The key property that only counterpropagating Alfv\'en waves 
interact is manifest here. For the slow-wave variables, 
we may introduce generalised Elsasser fields: 
\bea
\zpar^\pm = \upar\pm{\dBpar\over \sqrt{4\pi\rho_0}}\lt(1+{v_A^2\over c_s^2}\rt)^{1/2}.
\eea
Straightforwardly, the evolution equation for these fields is
\bea
\nonumber
{\dd\zpar^\pm\over \dd t} \mp {v_A\over \sqrt{1+v_A^2/c_s^2}}\dpar\zpar^\pm
&=& -{1\over 2}\lt(1\mp{1\over \sqrt{1+v_A^2/c_s^2}}\rt)\bl\{\zeta^+,\zpar^\pm\br\}\\
&&  -{1\over 2}\lt(1\pm{1\over \sqrt{1+v_A^2/c_s^2}}\rt)\bl\{\zeta^-,\zpar^\pm\br\}.
\label{eq_zpar}
\eea
This equation reduces to the parallel part of \eeqref{eq_z} in the limit 
$v_A\ll c_s$. This is known as the high-$\beta$ limit, with the
plasma beta defined by $\beta=8\pi p_0/B_0^2=(2/\gamma)c_s^2/v_A^2$. 
We see that only in this limit do the slow 
waves interact exclusively with the counterpropagating Alfv\'en waves. 
For general $\beta$, the phase speed of the slow waves is smaller than 
that of the Alfv\'en waves and, therefore, Alfv\'en waves can 
``catch up'' and interact with the slow waves that travel in the same 
direction. 

In astrophysical turbulence, $\beta$ tends to be moderately high:\footnote{The solar 
corona, where $\beta\sim10^{-6}$, is one prominent exception.} 
for example, in the interstellar medium, $\beta\sim10$ by order of magnitude. 
In the high-$\beta$ limit, which is equivalent to the incompressible 
approximation for the slow waves, density fluctuations 
are due solely to the entropy mode. They 
decouple from the slow-wave cascade 
and are passively mixed by the Alfv\'en-wave turbulence: $d\drho/dt=0$ 
[\eeqref{eq_ds} or \eeqref{eq_drho}, $c_s\gg v_A$]. 
By a dimensional argument similar to K41, 
the spectrum of such a field is expected to follow the spectrum 
of the underlying turbulence \cite{Obukhov}: in the GS theory, 
$\kperp^{-5/3}$. It is precisely 
the electron-density spectrum (deduced from observations of the scintillation 
of radio sources due to the scattering of radio waves by the 
interstellar medium) that provides the evidence of the $k^{-5/3}$ scaling 
in the interstellar turbulence \cite{Armstrong_Rickett_Spangler}. 
The explanation of this density spectrum in terms of passive mixing 
of the entropy mode, originally conjectured by Higdon \cite{Higdon1}, 
is developed on the basis of the GS theory in Ref.\ \cite{Lithwick_Goldreich01}.

Thus, the anisotropy and critical balance taken as ordering assumptions lead to a 
neat decomposition of the MHD turbulent cascade into a decoupled 
Alfv\'en-wave cascade 
and cascades of slow waves and entropy fluctuations 
passively scattered/mixed by the Alfv\'en waves.\footnote{Eqs.\ \exref{eq_ds}, 
\exref{eq_zeta} and \exref{eq_zpar} imply that, at 
arbitrary $\beta$, there are five conserved quantities: 
$\Es=\la|\ds|^2\ra$ (entropy fluctuations), 
$\Eperp^\pm=\la|\vdel\zeta^\pm|^2\ra$ (right/left-propagating Alfv\'en waves), 
$\Epar^\pm=\la|\zpar^\pm|^2\ra$ (right/left-propagating slow waves). 
$\Eperp^+$ and $\Eperp^-$ are always cascaded by interaction with each 
other, $\Es$ is passively mixed by $\Eperp^+$ and $\Eperp^-$, 
$\Epar^\pm$ are passively scattered by $\Eperp^\mp$ and, unless 
$\beta\gg 1$, also by $\Eperp^\pm$.} 
The validity of this decomposition and, especially, 
of the RMHD equations \exsandref{RMHD_psi}{RMHD_phi} turns out to 
extend to collisionless scales, where the MHD equations \exsref{MHD_rho}{MHD_B}
cannot be used: this will be briefly discussed in \secref{sec_GK}. 
}

\begin{figure}[t]
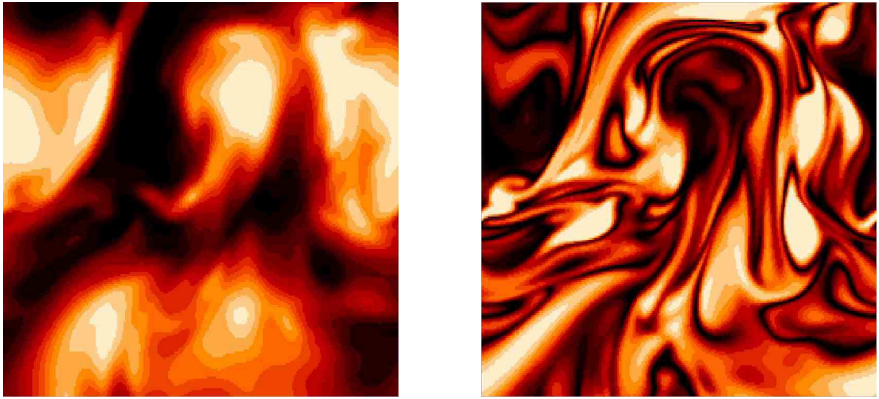

\centering 
\includegraphics[height=5.25cm]{u_z480_square.epsf} 
\hskip1cm
\includegraphics[height=5.25cm]{B_z480_square.epsf} 
\caption{Cross sections of the absolute values of $\vu$ (left panel) 
and $\vB$ (right panel) in the saturated state of a simulation 
with $\Rey\simeq 100$, $\Pm=10$ (run B of Ref.\ \cite{SCTMM_stokes}).}
\label{fig_slices}
\end{figure}

\section{Isotropic MHD turbulence} 
\label{sec_iso}

Let us now consider the case of isotropic MHD turbulence, i.e., 
a turbulent plasma where no mean field is imposed externally. 
Can the scaling theories reviewed above be adapted to this case?
A popular view, due originally to Kraichnan \cite{Kraichnan}, 
is that everything remains the same with 
the role of the mean field $\vB_0$ now played 
by the magnetic fluctuations at the outer scale, $\dB_L$, 
while at smaller scales, the turbulence is again a local (in scale) 
cascade of Alfv\'enic ($\du_l\sim\dB_l$) fluctuations.
This picture is only plausible if 
the magnetic energy is dominated by the outer-scale 
fluctuations, an assumption that does not appear to hold 
in the numerical simulations of forced isotropic MHD turbulence 
\cite{SCTMM_stokes}. Instead, the magnetic energy is 
concentrated at small scales, where the magnetic fluctuations 
significantly exceed the velocity fluctuations, with no sign 
of the scale-by-scale equipartition implied for an Alfv\'enic 
cascade.\footnote{This is true for the case of {\em forced} turbulence. 
Simulations of the decaying case \cite{Biskamp_Mueller} 
present a rather different picture: there is still
no scale-by-scale equipartition but the magnetic energy 
heavily dominates at the {\em large} scales --- most likely due 
to a large-scale force-free component controlling the decay. 
The difference between the numerical results on the decaying 
and forced MHD turbulence points to another break down in 
universality in stark contrast with the 
basic similarity of the two regimes in the hydrodynamic case.} 
These features are 
especially pronounced when the magnetic Prandtl number 
$\Pm=\nu/\eta=\Rm/\Rey\gg 1$, i.e., when the magnetic 
cutoff scale lies below the viscous cutoff of the velocity 
fluctuations (\figref{fig_slices}). The numerically more 
accessible case of $\Pm\gtrsim1$, while non-asymptotic and, therefore, 
harder to interpret, retains most of the features of 
the large-$\Pm$ regime. 
A handy formula for $\Pm$ based on the Spitzer \cite{Spitzer_book}
values of $\nu$ and $\eta$ for fully ionised plasmas is 
\bea
\Pm\sim10^{-5}T^4/n,
\label{Pm_formula}
\eea 
where $T$ is the temperature in Kelvin and $n$ the particle density in cm$^{-3}$. 
\Eqref{Pm_formula} tends to give very large values for hot diffuse 
astrophysical plasmas: e.g., $10^{11}$ for the warm interstellar medium, 
$10^{29}$ for galaxy clusters. 

Let us examine the situation in more detail. 
In the absence of a mean field, all magnetic fields are generated 
and maintained by the turbulence itself, i.e., 
isotropic MHD turbulence is the saturated state of the turbulent 
(small-scale) dynamo. Therefore, we start by considering 
how a weak (dynamically unimportant) magnetic field is amplified 
by turbulence in a large-$\Pm$ MHD fluid 
and what kind of field can be produced this way. 

\subsection{Small-scale dynamo}
\label{sec_dynamo}

Many specific deterministic flows 
have been studied numerically and analy\-tically and shown to 
be dynamos \cite{Childress_Gilbert}. 
While rigourously determining whether any given flow is a dynamo 
is virtually always a formidable mathematical challenge, 
the combination of numerical and analytical experience of 
the last fifty years suggests that smooth three-dimensional flows 
with chaotic trajectories tend to have the dynamo property 
provided the magnetic Reynolds number exceeds a certain 
threshold, $\Rm>\Rmc\sim10^1...10^2$. 
In particular, the ability of Kolmogorov turbulence to amplify 
magnetic fields is a solid numerical fact first established 
by Meneguzzi {\em et al.}\ in 1981 \cite{Meneguzzi_Frisch_Pouquet} 
and since then confirmed in many numerical studies with ever-increasing 
resolutions (most recently \cite{SCTMM_stokes,Haugen_Brandenburg_Dobler}). 
It was, in fact, Batchelor who realised already in 1950 \cite{Batchelor} 
that the growth of magnetic fluctuations in a random flow 
should occur simply as a consequence of the random stretching 
of the field lines and that it should proceed at the rate of strain 
associated with the flow. In Kolmogorov turbulence, 
the largest rate of strain $\sim\du_l/l$ is associated with the smallest 
scale $l\sim\ld$ --- the viscous scale, so it is the viscous-scale 
motions that dominantly amplify the field (at large $\Pm$). Note that the 
velocity field at the viscous scale is random but smooth, so the small-scale 
dynamo in Kolmogorov turbulence belongs to the same class as 
fast dynamos in smooth single-scale flows \cite{Childress_Gilbert,SCTMM_stokes}. 

\begin{figure}[t]
\centering 
\includegraphics[height=3.25cm]{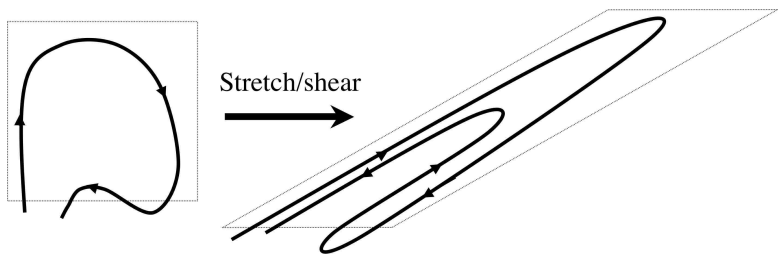} %
\caption{Stretching/shearing a magnetic-field line.}
\label{fig_stretch}
\end{figure}

A repeated application of random stretching/shearing to a tangled 
magnetic field produces direction reversals at arbitrarily small scales, 
giving rise to a folded field structure (\figref{fig_stretch}). 
It is an essential property of this structure that the strength of the 
field and the curvature of the field lines are anticorrelated: 
wherever the field is growing it is relatively straight (i.e., curved only on 
the scale of the flow), whereas in the bending regions, where 
the curvature is large, the field is weak. 
A quantitative theory of the folded structure 
can be constructed based on the joint statistics of the field 
strength $B=|\vB|$ and curvature $\vK=\vb\cdot\vdel\vb$, 
where $\vb=\vB/B$
\cite{SCMM_folding}.\footnote{This is not the only existing 
way of diagnosing the field structure. Ott and coworkers studied 
field reversals by measuring magnetic-flux cancellations \cite{Ott_review}. 
Chertkov {\em et al.} \cite{Chertkov_etal_dynamo} 
considered two-point correlation functions of the magnetic field 
in a model of small-scale dynamo and found large-scale correlations 
along the field and short-scale correlations across.} 
The curvature is a quantity easily measured 
in numerical simulations, which confirm the overall straightness 
of the field and the curvature--field-strength anticorrelation 
\cite{SCTMM_stokes}. At the end of this section, we shall give 
a simple demonstration of the validity of the folded structure. 

The scale of the direction reversals is limited from below only by 
Ohmic diffusion: for Kolmogorov turbulence, balancing the rate of 
strain at the viscous scale with diffusion and taking $\Pm\gg 1$ gives 
the resistive cutoff $\lres$: 
\bea
\du_{\ld}/\ld\sim \eta/\lres^2 \quad\Rightarrow\quad
\lres\sim\Pm^{-1/2}\ld.
\label{lres_kin}
\eea 
%Thus, the folded structure, which essentially required 
%a scale separation between the velocity (viscous)
%cutoff at $\ld$ and scale of direction reversals, 
%is a feature of the large-$\Pm$ regime. 

If random stretching gives rise to magnetic fields with 
reversals at the resistive scale, why are these fields not 
eliminated by diffusion? In other words, how is the small-scale dynamo
possible? It turns out that, in 3D, there are magnetic-field 
configurations that can be stretched without being destroyed 
by the concurrent refinement of the reversal scale and that 
add up to give rise to exponential growth of the magnetic energy. 
Below we give an analytical demonstration of this. 
It is a (somewhat modified) version of an ingenious argument originally 
proposed by Zeldovich {\em et al.}\ in 1984 \cite{Zeldovich_etal_linear}. 
A reader looking solely for a broad qualitative 
picture of isotropic MHD turbulence may skip to \secref{sec_sat}.\\ 

\begin{figure}[t]
\centering 
\includegraphics[height=3cm]{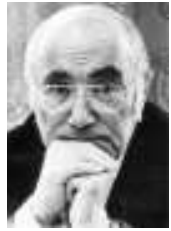} %
%http://scienceworld.wolfram.com/biography/Zeldovich.html
\caption{Ya.\ B.\ Zeldovich (1914-1987) (photo courtesy of M.\ Ya.\ Ovchinnikova).}
\label{fig_KZ}
\end{figure}

{\small

\noindent
{\bf Overcoming diffusion.} Let us study magnetic fields with reversals 
at subviscous scales: at these scales, the velocity 
field is smooth and can, therefore, be expanded
\bea
\label{u_lin}
u^i(t,\vx) = u^i(t,{\bf 0}) + \sigma^i_m(t)x^m + \dots,
\eea 
where $\sigma^i_m(t)$ is the rate-of-strain tensor. 
The expansion is around some reference point $\vx={\bf 0}$. We can always go 
to the reference frame that moves with the velocity at this point, so that 
$u^i(t,{\bf 0})=0$. 
Let us seek the solution to \eeqref{ind_eq} with velocity \exref{u_lin} 
as a sum of random plane waves with time-dependent wave vectors: 
\bea
\label{zeld_ansatz}
B^i(t,\vx) = \int{d^3 k_0\over (2\pi)^3}\,\tB^i(t,\vk_0)e^{i\tvk(t,\vk_0)\cdot\vx},
\eea 
where $\tvk(0,\vk_0)=\vk_0$, so $\tB^i(0,\vk_0)=B_0^i(\vk_0)$ is the Fourier 
transform of the initial field. 
Since \eeqref{ind_eq} is linear, it is sufficient to ensure 
that each of the plane waves is individually a solution. 
This leads to two {\em ordinary} differential equations for every $\vk_0$: 
\bea
\dd_t\tB^i = \sigma^i_m\tB^m - \eta\tk^2\tB^i,\qquad
\dd_t\tk_l = -\sigma^i_l\tk_i,
\eea
subject to initial conditions $\tB^i(0,\vk_0) = B^i_0(\vk_0)$ 
and $\tk_l(0,\vk_0) = k_{0l}$.
The solution of these equations can be written explicitly in terms of 
the Lagragian transformation of variables $\vx_0\to\vx(t,\vx_0)$, where 
\bea
\dd_t x^i(t,\vx_0) = u^i(t,\vx(t,\vx_0)) = \sigma^i_m(t) x^m(t,\vx_0),\qquad
x^i(0,\vx_0)=x^i_0.
\eea
Because of the linearity of the velocity field, 
the strain tensor $\dd x^i/\dd x_0^m$ and its inverse $\dd x_0^r/\dd x^l$ 
are functions of time only. At $t=0$, they are unit matrices. 
At $t>0$, they satisfy
\bea
\dd_t{\dd x^i\over \dd x_0^m} = \sigma^i_l{\dd x^l\over \dd x_0^m},
\qquad
\dd_t{\dd x_0^r\over \dd x^l} = -\sigma^i_l{\dd x^r\over \dd x_0^i}.
\eea
We can check by direct substitution that 
\bea
\tB^i(t,\vk_0) = {\dd x^i\over \dd x_0^m}\,B_0^m(\vk_0)\exp\lt[-\eta\int_0^t dt'\tk^2(t')\rt],\quad
\tk_l(t,\vk_0) = {\dd x_0^r\over \dd x^l}\,k_{0r}.
\label{k_sln}
\eea
These formulae express the evolution of one mode in 
the integral \exref{zeld_ansatz}. Using the fact that 
$\det(\dd x_0^r/\dd x^l)=1$ in an incompressible flow 
and that \eeqref{k_sln} therefore 
establishes a one-to-one correspondence $\vk\leftrightarrow\vk_0$, 
it is easy to prove that the volume-integrated magnetic energy is the 
sum of the energies of individual modes:
\bea
\label{Bsq_total}
\Bsq(t) \equiv \int d^3x\,|\vB(t,\vx)|^2 = 
\int{d^3 k_0\over (2\pi)^3}\,|\tvB(t,\vk_0)|^2.
\eea
From \eeqref{k_sln}, 
\bea
|\tvB(t,\vk_0)|^2 = \vB_0(\vk_0)\cdot\vM(t)\cdot\vB_0^*(\vk_0)
\exp\lt[-2\eta\int_0^t dt'\,\vk_0\cdot\vM^{-1}(t')\cdot\vk_0\rt],
\label{Bsq_sln}
\eea
where the matrices $\vM$ and $\vM^{-1}$ have elements defined by 
\bea
M_{mn}(t) = {\dd x^i\over \dd x_0^m}{\dd x^i\over \dd x_0^n}\quad\mathrm{and}\quad 
M^{rs}(t) = {\dd x_0^r\over \dd x^l}{\dd x_0^s\over \dd x^l}, 
\eea
respectively. 
They are the co- and contravariant metric tensors of the inverse Lagrangian 
transformation $\vx\to\vx_0$. 

Let us consider the simplest possible case of a flow \exref{u_lin} with 
constant $\hat\sigma=\diag\lt\{\lambda_1,\lambda_2,\lambda_3\rt\}$, 
where $\lambda_1>\lambda_2\ge0>\lambda_3$ and $\lambda_1+\lambda_2+\lambda_3=0$ 
by incompressibility. 
Then $\vM=\diag\lt\{e^{2\lambda_1 t},e^{2\lambda_2 t},e^{2\lambda_3 t}\rt\}$ 
and \eeqref{Bsq_sln} becomes, in the limit $t\to\infty$, 
\bea
|\tvB(t,\vk_0)|^2 \sim |B_{0}^1(\vk_0)|^2\exp\lt[2\lambda_1t-\eta
\lt({k_{01}^2\over \lambda_1} + {k_{02}^2\over \lambda_2} 
+ {k_{03}^2\over |\lambda_3|}\,e^{2|\lambda_3|t}\rt)\rt],
\label{B_estimate}
\eea
where we have dropped terms that decay exponentially with time 
compared to those retained.\footnote{ 
If $\lambda_2=0$, $k_{02}^2/\lambda_2$ in \eeqref{B_estimate}
is replaced with $2k_{02}^2 t$. The case $\lambda_2<0$ is treated 
in a way similar to that described below and also leads to magnetic-energy 
growth. The difference is that for $\lambda_2\ge0$, the magnetic structures 
are flux sheets, or ribbons, while for $\lambda_2<0$, they are flux ropes.} 
We see that for most $\vk_0$, the corresponding modes decay superexponentially 
fast with time. The domain in the $\vk_0$ space containing modes that are not exponentially 
small at any given time $t$ is given by 
\bea
{k_{01}^2\over \lambda_1^2t/\eta} + {k_{02}^2\over \lambda_1\lambda_2t/\eta} 
+ {k_{03}^2\over \lambda_1|\lambda_3|t e^{2\lambda_3t}/\eta} <\const.
\eea
The volume of this domain at time $t$ is 
$\sim\lambda_1^2(\lambda_2|\lambda_3|)^{1/2} (t/\eta)^{3/2}e^{\lambda_3t}$. 
Within this volume, $|\tvB(t,\vk_0)|^2 \sim |B_0^1(\vk_0)|^2e^{2\lambda_1t}$. 
Using $\lambda_3=-\lambda_1-\lambda_2$ and \eeqref{Bsq_total}, we get
\bea
\Bsq(t)\propto\exp\lt[(\lambda_1-\lambda_2)t\rt].
\label{Bsq_growth}
\eea

Let us discuss the physics behind the Zeldovich {\em et al.}\ calculation sketched 
above. When the magnetic field is stretched by the flow, it naturally aligns with 
the stretching Lyapunov direction: $\vB\sim\ve_1 B_0^1e^{\lambda_1 t}$. 
The wave vector $\vk$ has a tendency to align with the compression direction: 
$\vk\sim\ve_3 k_{03} e^{|\lambda_3|t}$, which makes most modes decay 
superexponentially. The only ones that survive are those whose 
$\vk_0$'s were nearly perpendicular to $\ve_3$, with the permitted angular deviation 
from $90^\circ$ decaying exponentially in time $\sim e^{-|\lambda_3|t}$. 
Since the magnetic field is solenoidal, $\vB_0\perp\vk_0$, the modes 
that get stretched the most have $\vB_0\parallel\ve_1$ and 
$\vk_0\parallel\ve_2$ (\figref{fig_zeldovich}). In contrast, 
in 2D, the field aligns with $\ve_1$ and must, therefore, 
reverse along $\ve_2$, which is always the compression direction 
(\figref{fig_zeldovich}), so 
the stretching is always overwhelmed by the diffusion and no dynamo is 
possible (as should be the case according to the rigourous 
early result of Zeldovich \cite{Zeldovich_antidynamo}). 

\begin{figure}[t]
\centering 
\includegraphics[height=5cm]{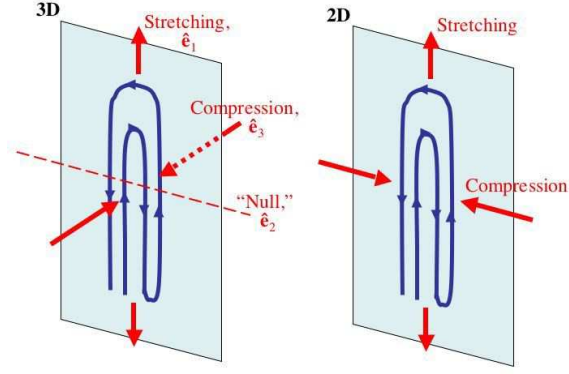} %
\caption{Magnetic fields vs.\ the Lyapunov directions (from \cite{SCTMM_stokes}).
Zeldovich {\em et al.}\ \cite{Zeldovich_etal_linear} did not give this exact 
interpretation of their calculation because the folded structure of the field was 
not yet clearly understood at the time.}
\label{fig_zeldovich}
\end{figure}

The above construction can be generalised to time-dependent and random 
velocity fields. The matrix $\vM$ is symmetric and can, therefore, 
be diagonalised by an appropriate rotation $\vR$ of the coordinate system: 
$\vM=\vR^T\cdot\vL\cdot\vR$, where, by definition, 
$\vL=\diag\lt\{e^{\zeta_1(t)},e^{\zeta_2(t)},e^{\zeta_3(t)}\rt\}$. 
It is possible to prove that, as $t\to\infty$, $\vR(t)\to\lt\{\ve_1,\ve_2,\ve_3\rt\}$ 
and $\zeta_i(t)/2t\to\lambda_i$, where $\ve_i$ are constant orthogonal 
unit vectors, which make up the Lyapunov basis, and $\lambda_i$ are the Lyapunov 
exponents of the flow \cite{Goldhirsch_Sulem_Orszag}. 
The instantaneous values of $\zeta_i(t)/2t$ are called 
finite-time Lyapunov exponents. For a random flow, $\zeta_i(t)$ are random functions. 
\Eqref{Bsq_growth} generalises to 
\bea
\la\overline{B^2}\ra(t)\propto\overline{\exp\lt[(\zeta_1-\zeta_2)/2\rt]},
\label{Bsq_zetas}
\eea 
where the overline means 
averaging 
over the distribution of $\zeta_i$. 
The only random flow for which this distribution is known is 
a Gaussian white-in-time velocity first considered in the dynamo context in 
1967 by Kazantsev \cite{Kazantsev}.\footnote{As the only analytically solvable 
model of random advection, Kazantsev's model has played a crucial role. 
Developed extensively in 1980s by Zeldovich and coworkers \cite{Almighty_Chance}, 
the model became a tool of choice in the theories of anomalous scaling 
and intermittency that flourished in 1990s \cite{Falkovich_Gawedzki_Vergassola} 
(in this context, it has been associated with the name of Kraichnan who, independently 
from Kazantsev, proposed to use it for the passive scalar problem \cite{Kraichnan_model}). 
It remains useful to this day as old theories are reevaluated and new questions 
demand analytical answers \cite{SCMM_folding}.}
%A linear Kazantsev velocity has a Gaussian rate-of-strain tensor that satisfies
%\bea
%\la\sigma^i_m(t)\sigma^j_n(t')\ra = {4\over 3}\,\lambda_1\delta(t-t') 
%\lt[\delta^{ij}\delta_{mn} - {1\over 4}\lt(\delta^i_m\delta^j_n 
%+\delta^i_n\delta^j_m\rt)\rt].
%\eea
The distribution of $\zeta_i$ for this flow is Gaussian in the 
long-time limit and \eeqref{Bsq_zetas} gives $\Bsq\propto e^{(5/4)\lambda_1 t}$, 
where $\lambda_1=\la\zeta_1\ra/2t$ \cite{Chertkov_etal_dynamo}. 
For Kazantsev's velocity, it is also possible to calculate 
the magnetic-energy spectrum \cite{Kazantsev,Kulsrud_Anderson}, 
which is the spectrum of the direction reversals. 
It has a peak at the resistive scale and a $k^{+3/2}$ power law 
stretching across the subviscous range, $\ld^{-1}\ll k\ll\lres^{-1}$. 
This scaling appears to be corroborated by numerical 
simulations \cite{SCTMM_stokes,Haugen_Brandenburg_Dobler}.\\ 

\noindent 
{\bf Folded structure revisited.} We shall now give a very simple 
demonstration that linear 
stretching does indeed produce folded fields with straight/curved 
field lines corresponding to larger/smaller field strength. 
Using \eeqref{ind_eq}, we can write 
evolution equations for the field strength $B=|\vB|$, 
the field direction $\vb=\vB/B$ 
and the field-line curvature $\vK=\vb\cdot\vdel\vb$. 
Omitting the resistive terms, 
\bea
\label{eq_B}
{dB\over dt} &=& \bl(\vb\vb:\vdel\vu\br)B,\\
\label{eq_b}
{d\vb\over dt} &=& \vb\cdot\bl(\vdel\vu\br)\cdot\bl(\unity-\vb\vb\br),\\ 
\nonumber
{d\vK\over dt} &=& \vK\cdot\bl(\vdel\vu\br)\cdot\bl(\unity-\vb\vb\br) 
- 2(\vb\vb:\vdel\vu)\vK - \bl[\vb\cdot\bl(\vdel\vu\br)\cdot\vK\br]\vb\\
&& + \vb\vb:\bl(\vdel\vdel\vu\br)\cdot\bl(\unity-\vb\vb\br).
\label{eq_K}
\eea
For simplicity, we again use the velocity field \exref{u_lin} with 
constant $\hat\sigma=\diag\lt\{\lambda_1,\lambda_2,\lambda_3\rt\}$. 
Then the stable fixed point of \eeqref{eq_b} in the comoving frame 
is $\vb=\ve_1$ (magnetic field aligns with the principal stretching direction), 
whence $B\propto e^{\lambda_1 t}$. Since $\vK\cdot\vb=0$, we set $K_1=0$. 
Denoting $\vW=\vb\vb:\vdel\vdel\vu$, we can now write \eeqref{eq_K} as
\bea
\label{eq_K23}
{d K_2\over dt} = -\bl(2\lambda_1-\lambda_2\br)K_2 + \Sigma_2,\qquad
{d K_3\over dt} = -\bl(3\lambda_1+\lambda_2\br)K_3 + \Sigma_3.
\eea
Both components of $\vK$ decay exponentially,\footnote{$K_3$ decays 
faster than $K_2$. If the velocity is exactly linear ($\vW=0$), $\vK$ 
aligns with $\ve_2$ and decreases indefinitely, while 
the combination $BK^{1/(2-\lambda_2/\lambda_1)}$ stays 
constant. This rhymes with the result that can be proven 
for a linear Kazantsev velocity: 
at zero $\eta$, $B\propto e^{\zeta_1/2}$, while 
$BK^{1/2}\propto e^{\zeta_2/4}$ and $\lambda_2=\la\zeta_2\ra/2t=0$.} 
until they are comparable to the inverse scale of the velocity field 
(i.e., the terms containing $\vdel\vdel\vu$ become important). 
The stationary solution is
$K_2=\Sigma_2/(2\lambda_1-\lambda_2)$, 
$K_3=\Sigma_3/(3\lambda_1+\lambda_2)$. 

If the field is to reverse direction, it must turn somewhere 
(see \figref{fig_stretch}). At such a turning point, the field must be perpendicular 
to the stretching direction. Setting $b_1=0$, we find two fixed points 
of \eeqref{eq_b} under this condition: $\vb=\ve_2$ and $\vb=\ve_3$. 
Only the former is stable, so the field at the turning point will tend 
to align with the ``null'' direction. Thus, 
stretching favours configurations with field reversals along 
the ``null'' direction, which are also those that survive diffusion
(see \figref{fig_zeldovich}). From \eeqref{eq_K} we find that 
at the turning point, $K_2=0$, $K_3=\Sigma_3/(\lambda_1+3\lambda_2)$, while 
$K_1$ grows at the rate $\lambda_1-2\lambda_2$ (assumed positive). 
This growth continues until limited by diffusion at $K\sim1/\lres$. 
The strength of the field in this curved region is $\propto e^{\lambda_2t}$, 
so the fields are weaker than in the straight segments, where $B\propto e^{\lambda_1t}$. 

When the problem is solved for the Kazantsev velocity, 
the above solution generalises to a field of random curvatures 
anticorrelated with the magnetic-field strength and with a 
stationary PDF of $K$ that has a
peak at $K\sim\mathrm{flow\ scale}^{-1}$ and a power tail $\sim K^{-13/7}$ 
describing the distribution of curvatures at the turning points 
\cite{SCMM_folding}. Numerical simulations support these results \cite{SCTMM_stokes}. 
} %end of small font

\begin{figure}[t]
\centering 
\begin{tabular}{ccccc}
\includegraphics[height=3cm]{batchelor.epsf} & &  
\includegraphics[height=3cm]{schlueter.epsf} & & 
\includegraphics[height=3cm]{biermann.epsf}\\ 
G.\ K.\ Batchelor (1920-2000) & &
%http://www-history.mcs.st-andrews.ac.uk/history/PictDisplay/Batchelor.html
A.\ Schl\"uter & & L.\ Biermann (1907-1986)  
%http://www.aip.org/history/esva/catalog/esva/Biermann_Ludwig.html
\end{tabular}
\caption{Photo of A.\ Schl\"uter courtesy of MPI f\"ur Plasmaphysik; 
photo of L.\ Biermann courtesy of Max-Planck-Gesellschaft/AIP Emilio Segr\`e Visual Archives.} 
\label{fig_BSB}
\end{figure}

\subsection{Saturation of the dynamo}
\label{sec_sat}

The small-scale dynamo gave us exponentially growing magnetic fields 
with energy concentrated at small (resistive) scales. 
How is the growth of magnetic energy saturated and what is the final state? 
Will magnetic energy stay at small scales or will it proceed to scale-by-scale 
equipartition via some form of inverse cascade? 
This basic dichotomy dates back to the 1950 papers by Batchelor \cite{Batchelor} 
and Schl\"uter \& Biermann \cite{Schlueter_Biermann}. 
Batchelor thought that magnetic field was basically analogous to 
the vorticity field $\vo=\vdel\times\vu$ (which satisfies 
the same equation \exref{ind_eq} except for the difference between $\eta$ and $\nu$) 
and would, therefore, saturate at a low energy, $\Bsq\sim\Rey^{-1/2}\usq$, 
with a spectrum peaked at the viscous scale. 
Schl\"uter and Biermann disagreed and argued that the saturated state would be 
a scale-by-scale balance between the Lorentz and inertial forces, 
with turbulent motions at each scale giving rise to magnetic 
fluctuations of matching energy at the same scale. 
Schl\"uter and Biermann's argument (and, implicitly, also 
Batchelor's) was based on the assumption of {\bf locality of interaction} 
(in scale space) between the magnetic and velocity fields: locality both of the dynamo 
action and of the back reaction. As we saw in the previous section, 
this assumption is certainly incorrect for the dynamo: a linear velocity field, 
i.e., a velocity field of a formally infinitely large (in practice, viscous) 
scale can produce magnetic fields with reversals at the smallest scale 
allowed by diffusion. A key implication of the folded structure of these 
fields concerns the Lorentz force, the essential part of which, in the case of 
incompressible flow, is the curvature force $B^2\vK$. Since it is a quantity 
that depends only on the parallel gradient of the magnetic field and does 
not know about direction reversals, it will possess a degree of velocity-scale 
spatial coherence necessary to oppose stretching. Thus, a field that is 
formally at the resistive scale will exert a back reaction at the scale 
of the velocity field. In other words, {\bf interactions are nonlocal:} 
a random flow at a given scale $l$, 
having amplified the magnetic fields at the resistive scale $\lres\ll l$, 
will see these magnetic fields back react at the scale $l$. 
Given the nonlocality of back reaction, we can update 
Batchelor's and Schl\"uter \& Biermann's scenarios for saturation 
in the following way \cite{SCHMM_ssim,SCTMM_stokes}. 

The magnetic energy is amplified by the viscous-scale motions 
until the field is strong enough to resist stretching, i.e., until
$\vB\cdot\vdel\vB\sim\vu\cdot\vdel\vu\sim \du_{\ld}^2/\ld$. 
Since $\vB\cdot\vdel\vB\sim B^2K\sim B^2/\ld$ (folded field), 
this happens when 
\bea
\Bsq\sim \du_{\ld}^2\sim\Rey^{-1/2}\usq. 
\label{Bsq_Batchelor}
\eea
Let us suppose that the viscous motions are suppressed 
by the back reaction, at least in their ability 
to amplify the field. Then the motions at larger scales in 
the inertial range come into play: while their rates of strain 
and, therefore, the associated stretching rates are smaller 
than that of the viscous-scale motions, 
they are more energetic [see \eeqref{E_K41}], 
so the magnetic field is too weak to resist being stretched by them. 
As the field continues to grow, it will suppress the motions 
at ever larger scales. If we define a stretching scale $\ls(t)$ 
as the scale of the motions whose energy is 
$\du_{\ls}\sim\Bsq(t)$, we can estimate
\bea
{d\over dt}\Bsq\sim {\du_{\ls}\over \ls}\Bsq\sim{\du_{\ls}^3\over \ls}
\sim\epsilon = \const\quad\Rightarrow\quad
\Bsq(t)\sim\epsilon t.
\label{Bsq_t}
\eea
Thus, exponential growth gives way to secular 
growth of the magnetic energy. This is accompanied by elongation of the folds 
(their length is always of the order of the stretching scale, 
$\lpar\sim\ls$), while the resistive (reversal) scale increases because 
the stretching rate goes down: 
\bea
\lpar(t)\sim\ls(t)\sim\du_{\ls}^3/\epsilon\sim\sqrt{\epsilon}\,t^{3/2},
\quad
\lres(t)\sim\lt[\eta/(\du_{\ls}/\ls)\rt]^{1/2}\sim\sqrt{\eta t}.
\label{ll_t}
\eea
%Since $\lpar$ grows faster than $\lres$, the ``aspect ratio'' of 
%the folds increases. 
This secular stage can continue until the entire inertial range is suppressed, 
$\ls\sim L$, at which point saturation must occur. 
This happens after $t\sim \epsilon^{-1/3}L^{2/3}\sim L/\du_L$. 
Using \eqsref{Bsq_t}{ll_t}, we have, in saturation,
\bea
\Bsq\sim\usq,\quad\lpar\sim L,\quad
\lres\sim\lt[\eta/(\du_L/L)\rt]^{1/2}\sim\Rm^{-1/2}L.
\label{lres_sat}
\eea
Comparing \eqsandref{lres_sat}{lres_kin}, 
we see that the resistive scale has increased only 
by a factor of $\Rey^{1/4}$ over its value in the weak-field growth stage. 
Note that this imposes a very stringent 
requirement on any numerical experiment striving to distinguish between 
the viscous and resistive scales: $\Pm\gg \Rey^{1/2}\gg 1$. 

\begin{figure}[t]
\centering 
\includegraphics[height=3cm]{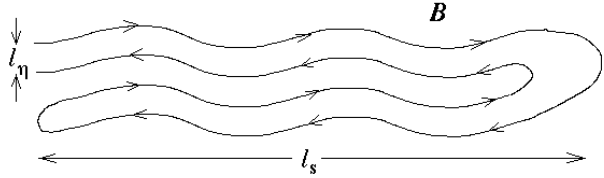} %
\caption{Alfv\'en waves propagating along folded fields (from \cite{SCHMM_ssim}).}
\label{fig_waves}
\end{figure}

If, as in the above scenario, the magnetic field retains its 
folded structure in saturation, with direction reversals 
at the resistive scale, this explains qualitatively why 
the numerical simulations of the developed isotropic MHD 
turbulence with $\Pm\ge1$ \cite{SCTMM_stokes} 
show the magnetic-energy pile-up at the small scales.
What then is the saturated state of the turbulent velocity field? 
We assumed above that the inertial-range motions were 
``suppressed'' --- this applied to their ability to amplify 
magnetic field, but needed not imply a complete evacuation 
of the inertial range. Indeed, simulations at modest $\Pm$ 
show a powerlike velocity spectrum 
\cite{SCTMM_stokes,Haugen_Brandenburg_Dobler}. 
The most obvious class of motions that can populate the inertial 
range without affecting the magnetic-field strength 
are a type of Alfv\'en waves that propagate not along a 
mean (or large-scale) magnetic field but along the folded 
structure (\figref{fig_waves}). Mathematically, the dispersion 
relation for such waves is derived 
via a linear theory carried out for the inertial-range 
perturbations ($L^{-1}\ll k\ll\ld^{-1}$) of the tensor $B_iB_j$ 
(cf.\ \cite{Gruzinov_Diamond96}). 
The unperturbed state is the average of this tensor 
over the subviscous scales: $\la B_i B_j\ra = \bb_i\bb_j\Bsq$, 
where $\Bsq$ is the total magnetic energy and the tensor 
$\bb_i\bb_j$ only varies at the outer scale $L$ [\eeqref{lres_sat}]. 
The resulting dispersion relation is $\omega=\pm|\vk\cdot\vb|\Bsq^{1/2}$ 
\cite{SCHMM_ssim}. The presence of these waves 
will not change the resistive-scale-dominated nature of the 
magnetic-energy spectrum, but should be manifest in the 
kinetic-energy spectrum. There is, at present, no theory of 
a cascade of such waves, although a line of argument similar 
to \secref{sec_aniso} might work, since 
it does not depend on the field having a specific direction. 
A numerical detection of these waves is also a challenge for the future.

What we have proposed above can be thought of as a modernised version 
of the Schl\"uter \& Biermann scenario, retaining the intermediate 
secular-growth stage and saturation with $\Bsq\sim\usq$, but 
not scale-by-scale equipartition. However, an alternative possibility, 
which is in a similar relationship to Batchelor's scenario, 
can also be envisioned. In \eqsref{Bsq_t}{ll_t}, 
the scale $\lres$ at which diffusion cuts off the small-scale magnetic 
fluctuations was assumed to be determined by the stretching rate $\du_{\ls}/\ls$. 
However, since the nonlinear suppression of the viscous-scale eddies 
only needs to eliminate motions with $\vb\vb:\vdel\vu\neq0$ 
[\eeqref{eq_B}], two-dimensional ``interchange'' motions 
(velocity gradients $\perp\vb$) are, in principle, allowed to survive at the 
viscous scale. These could ``two-dimensionally'' mix the direction-reversing 
magnetic fields at the rate $\du_{\ld}/\ld$ --- much faster than the 
unsuppressed larger-scale stretching can amplify the field, --- 
with the consequence 
that the resistive scale is pinned at the value given by \eeqref{lres_kin} 
and the field cannot grow above the Batchelor limit \exref{Bsq_Batchelor}. 
The mixing efficiency of the suppressed motions is the key 
to choosing between the two saturation scenarios. 
Numerical simulations \cite{SCTMM_stokes} 
corroborate the existence of an intermediate stage of 
slower-than-exponential growth accompanied by fold elongation 
and a modest increase of the resistive scale [\eeqref{ll_t}]. 
This tips the scales in favour of the first scenario, but, 
in view of limited resolutions, we hesitate to declare the matter 
definitively resolved. 

\subsection{Turbulence and magnetic fields in galaxy clusters} 
\label{sec_clusters}

The intracluster medium (hereafter, ICM) is a hot ($T\sim10^8$ K) diffuse 
($n\sim10^{-2}...10^{-3}$ cm$^{-3}$) fully ionised plasma, which accounts 
for most of the luminous matter in the Universe (note that it is not 
entirely dissimilar from the ionised phases of the interstellar medium: 
e.g., the so-called hot ISM). 
It is a natural astrophysical environment to which the large-$\Pm$ 
isotropic regime of MHD turbulence appears to be applicable: indeed,  
\eeqref{Pm_formula} gives $\Pm\sim10^{29}$. 

The ICM is believed to be in a state of 
turbulence driven by a variety of mechanisms: merger events, 
galactic and subcluster wakes, active galactic nuclei. 
One expects the outer scale $L\sim10^2...10^3$ kpc 
and the velocity dispersions $\du_L\sim10^2...10^3$ km/s 
(a fraction of the sound speed). Indirect observational evidence 
supporting the possibility of a turbulent 
ICM with roughly these parameters already exists (an apparently powerlike 
spectrum of pressure fluctuations found in the Coma cluster \cite{Schuecker_etal}, 
broadened abundance profiles in Perseus believed to be caused 
by turbulent diffusion \cite{Rebusco_etal}), 
and direct detection may be achieved in the near future 
\cite{Inogamov_Sunyaev}. However, there is as yet no consensus on whether 
turbulence, at least in the usual hydrodynamic sense, is a generic feature of 
clusters \cite{Fabian_etal_Perseus2}. The main difficulty 
is the very large values of 
the ICM viscosity obtained via the standard estimate $\nu\sim\vth\mfp$, 
where $\vth\sim10^3$ km/s is the ion thermal speed and $\mfp\sim1...10$ kpc 
is the ion mean free path. This gives $\Rey\sim10^2$ if not 
less, which makes the existence of a well-developed inertial range problematic. 
Postponing the problem of viscosity until \secref{sec_plasma}, 
we observe that the small-scale dynamo does not, in fact, require a turbulent 
velocity field in the sense of a broad inertial range: 
in the weak-field regime discussed in \secref{sec_dynamo}, 
the dynamo was controlled by the smooth 
single-scale random flow associated with the viscous-scale motions; 
in saturation, we argued in \secref{sec_sat} that 
the main effect was the direct nonlocal 
interaction between the outer-scale (random) motions 
and the magnetic field. Given the available menu of 
large-scale stirring mechanisms in clusters, it is likely that, 
whatever the value of $\Rey$, the velocity field is random.\footnote{Numerical 
simulations of the large-$\Pm$ regime at currently accessible 
resolutions also rely on a random forcing to produce ``turbulence'' 
with $\Rey\sim1...10^2$ \cite{SCTMM_stokes,Haugen_Brandenburg_Dobler}.} 

The cluster turbulence is certainly 
magnetic. The presence of magnetic fields was first demonstrated 
for the Coma cluster, for which Willson detected in 1970 
a diffuse synchrotron radio emission \cite{Willson} and 
Kim {\em et al.}\ in 1990 were able to estimate directly the 
magnetic-field strength and scale 
using the Faraday rotation measure (RM) data \cite{Kim_etal_Coma}. 
Such observations of magnetic fields in clusters have now become a vibrant 
area of astronomy (reviewed most recently in \cite{Govoni_Feretti_review}), 
usually reporting a field $B\sim1...10\ \mu$G at scales 
$\sim1...10$ kpc \cite{Clarke_Kronberg_Boehringer}.\footnote{Because of the availability of 
the RM maps from extended radio sources in clusters, it is possible to go beyond 
field-strength and scale estimates and construct magnetic-energy 
spectra with spatial resolution of $\sim0.1$ kpc \cite{Vogt_Ensslin2}.} 
All of this field is small-scale fluctuations: 
no appreciable mean component has been detected. The field is 
dynamically significant: the magnetic energy is less but not much less than 
the kinetic energy of the turbulent motions. 

Do clusters fit the theoretical expectations reviewed above? 
The magnetic-field scale seen in clusters is usually 
10 to 100 times smaller than the expected outer scale of turbulent motions 
and, indeed, is also smaller than the viscous scale based on $\Rey\sim10^2$. 
However, it is certainly far above the resistive scale, which turns out 
be $\lres\sim10^3...10^4$ km! Faced with these numbers, we must  
suspend the discussion of cluster physics and finally take account 
of the fact that astrophysical bodies are made of plasma, not of 
an MHD fluid. 

\section{Enter plasma physics} 
\label{sec_plasma}

\subsection{Braginskii viscosity}
\label{sec_brag}

In all of the above, we have used the MHD equations \exsandref{NSEq}{ind_eq} 
to develop turbulence theories supposed to be relevant for astrophysical 
plasmas. Historically, such has been the approach followed in most 
of the astrophysical literature. The philosophy underpinning 
this approach is again that of universality: the ``microphysics'' 
at and below the dissipation scale are not expected to matter 
for the fluid-like dynamics at larger scales. 
However, in considering the MHD turbulence 
with large $\Pm$, we saw that dissipation scales, determined by the 
values of the viscosity $\nu$ and magnetic diffusivity $\eta$, 
played a very prominent role: the growth of the small-scale magnetic 
fields was controlled by the turbulent rate of strain at the viscous 
scale and resulted in the magnetic energy piling up, in the form of 
direction-reversing folded fields, at the resistive scale --- both 
in the growth and saturation stages of the dynamo. 
It is then natural to revisit the question 
of whether the Laplacian diffusion terms in 
\eqsref{NSEq}{ind_eq} are a good description of 
the dissipation in astrophysical plasmas. 

The answer to this question is, of course, that they are not. 
A necessary assumption 
in the derivation of these terms is that the ion 
cyclotron frequency $\Omega_i=eB/m_ic$ exceeds the 
ion-ion collision frequency $\nui$ or, equivalently, the ion 
gyroradius $\rho_i=\vth/\Omega_i$ exceeds the mean free path 
$\mfp=\vth/\nui$. 
This is patently not the case in many astrophysical plasmas:
for example, in galaxy clusters, $\mfp\sim1...10$ kpc, while $\rho_i\sim10^4$ km. 
In such a weakly collisional magnetised plasma, the momentum 
equation \exref{NSEq} assumes the following form, valid at spatial 
scales $\gg \rho_i$ and at time scales $\gg \Omega_i^{-1}$, 
\bea
\label{u_eq_brag}
{d\vu\over dt} = -\vdel\lt(\pperp+{B^2\over 2}\rt) 
+ \vdel\cdot\bl[\vb\vb(\pperp-\ppar + B^2)\br] + \vf,
\eea
where $\pperp$ and $\ppar$ are plasma pressures perpendicular 
and parallel to the local direction of the magnetic field, respectively, 
and we have used $\vB\cdot\vdel\vB=\vdel\cdot(\vb\vb B^2)$. 
The evolution of the magnetic field is controlled by the electrons --- 
the field remains frozen into the flow and we may use 
\eeqref{ind_eq} with $\eta = 0$. 

If we are interested in subsonic motions, $\vdel(\pperp+B^2/2)$ in 
\eeqref{u_eq_brag} can be found from the incompressibility condition 
$\vdel\cdot\vu=0$ and the only quantity still to be determined 
is $\pperp-\ppar$. The proper way to compute it is by a 
rather lengthy kinetic calculation due to Braginskii 
\cite{Braginskii}, which cannot be repeated here. 
The result of this calculation can, however, be obtained in the 
following heuristic way \cite{SCKHS_brag}. 

The fundamental property of 
charged particles moving in a magnetic field is the conservation 
of the first adiabatic invariant $\mu=m_i\vperp^2/2B$.\footnote{It may 
be helpful to the reader to think of this property 
as the conservation of the angular momentum of a gyrating particle: 
$m_i\vperp\rho_i\propto m_i\vperp^2/B=2\mu$.} 
When $\mfp\gg \rho_i$, 
this conservation is only weakly broken by collisions. 
As long as $\mu$ is conserved, any change 
in $B$ must be accompanied by a proportional change in $\pperp$.
Thus, the emergence of the pressure anisotropy is a natural 
consequence of the changes in the magnetic-field strength and vice 
versa: indeed, summing up the first adiabatic invariants of all particles, 
we get $\pperp/B=\const$. Then 
\bea
{1\over \pperp}{d\pperp\over dt} 
= {1\over B}{dB\over dt} 
-\nui\,{\pperp-\ppar\over \pperp},
\label{dlnB_approx}
\eea
where the second term on the right-hand sight represents 
the collisional relaxation of the pressure anisotropy $\pperp-\ppar$ 
at the rate $\nui\sim\vth/\mfp$.\footnote{This is 
only valid if the characteristic parallel scales $\kpar^{-1}$ 
of all fields are larger than $\mfp$. In the collisionless regime, 
$\kpar\mfp\gg 1$, we may assume that the 
pressure anisotropy is relaxed in the time particles streaming 
along the field cover the distance $\kpar^{-1}$: this entails 
replacing $\nui$ in \eeqref{dlnB_approx} by $\kpar\vth$.} 
Using \eeqref{eq_B} for $B$ 
and balancing the terms in the rhs of \eeqref{dlnB_approx}, we get 
\bea
\pperp-\ppar = \nupar\,{1\over B}{dB\over dt} = \nupar\vb\vb:\vdel\vu, 
\label{Brag_visc}
\eea
where $\nupar\sim p/\nui\sim\vth\mfp$ is the ``parallel viscosity.'' 
This equation turns out to be exact \cite{Braginskii} 
up to numerical prefactors in the definition of $\nupar$. 

The energy conservation law based on \eqsandref{u_eq_brag}{ind_eq} is
\bea
{d\over dt}\lt({\usq\over 2} + {\Bsq\over 2}\rt) 
= \epsilon - \nupar\bl<|\vb\vb:\vdel\vu|^2\br> 
= \epsilon - \nupar\lt<\lt({1\over B}{dB\over dt}\rt)^2\rt>, 
\eea
where Ohmic diffusion has been omitted. 
Thus, the Braginskii viscosity only dissipates 
such velocity gradients that change the strength of the 
magnetic field. The motions that do not affect $B$ are allowed 
to exist in the subviscous scale range. 
In the weak-field regime, these motions take the form of plasma 
instabilities. When the magnetic field is strong, a cascade of 
shear-Alfv\'en waves can be set up below the viscous scale. 
Let us elaborate. 

\subsection{Plasma instabilities} 

The simplest way to see that the pressure anisotropy in \eeqref{u_eq_brag} 
leads to instabilities is as follows \cite{SCKHS_brag}. Imagine a ``fluid'' 
solution with $\vu$, $\pperp$, $\ppar$, $\vB$ changing on viscous time and 
spatial scales, $t\sim|\vdel\vu|^{-1}\sim\ld/\du_{\ld}$ and $l\sim\ld$. 
Would such a solution be stable with respect to fast ($\omega\gg |\vdel\vu|^{-1}$) 
small-scale ($k\gg \ld^{-1}$) perturbations? Linearising \eeqref{u_eq_brag} 
and denoting perturbations by $\delta$, we get 
\bea
\nonumber
-i\omega\dvu &=& -i\vk\lt(\delta\pperp + B\delta B\rt) 
+ \lt(\pperp-\ppar + B^2\rt)\delta\vK\\ 
&&+\,\, i\vb\,\kpar\lt[\delta\pperp-\delta\ppar - \lt(\pperp - \ppar - B^2\rt) 
{\delta B/B}\rt],
\label{du_eq}
\eea
where the perturbation of the field curvature is 
$\delta\vK=\kpar^2\dvuperp/i\omega$ [see \eeqref{eq_K}]. 
We see that regardless of the 
origin of the pressure anisotropy, the shear-Alfv\'en-polarised perturbations 
($\dvu\propto\vk\times\vb$) have the dispersion relation 
\bea
\omega = \pm\kpar\lt(\pperp-\ppar+B^2\rt)^{1/2}. 
\label{firehose}
\eea
When $\ppar-\pperp>B^2$, $\omega$ is purely imaginary and we have 
what is known as the firehose instability 
\cite{Rosenbluth,Chandrasekhar_Kaufman_Watson,Parker_firehose,Vedenov_Sagdeev}. 
The growth rate of the instability is $\propto\kpar$, which means 
that the fastest-growing perturbations will be at scales far below 
the viscous scale or, indeed, the mean free path. Therefore, adopting 
the Braginskii viscosity [\eeqref{Brag_visc}] exposes a fundamental 
problem with the use of the MHD approximation for fully ionised 
plasmas: the equations are ill posed wherever $\ppar-\pperp>B^2$. 
To take into account the instability and its impact on the large-scale 
dynamics, the fluid equations must be abandoned and a kinetic 
description adopted. A linear kinetic calculation shows that 
the instability growth rate peaks at $\kpar\rho_i\sim1$, so the 
fluctuations grow fastest at the ion gyroscale. 
While the firehose instability occurs in regions where the velocity field 
leads to a decrease in the magnetic-field strength [\eeqref{Brag_visc}], 
a kinetic calculation of the pressure perturbations in \eeqref{du_eq} 
shows that another instability, called the mirror mode \cite{Parker_firehose}, 
is triggered wherever the field increases ($\pperp>\ppar$). 
Its growth rate is also $\propto\kpar$ and peaks at the ion gyroscale. 

In weakly collisional astrophysical plasmas such as 
the ICM, the random motions produced 
by the large-scale stirring will stretch and fold magnetic fields,  
giving rise to regions both of increasing and decreasing field 
strength (\secref{sec_dynamo}). The instabilities should, therefore, 
be present in weak-field regions where $|\ppar-\pperp|>B^2$ 
and, since their growth rates are much larger than 
the fluid rates of strain, their growth and saturation should have 
a profound effect on the structure of the turbulence. 
A quantitative theory of what exactly happens is not as 
yet available, but one might plausibly expect that the fluctuations 
excited by the instabilities will lead to some effective renormalisation 
of both the viscosity and the magnetic diffusivity. A successful theory of 
turbulence in clusters requires a quantitative calculation of 
this effective transport. In particular, this should resolve the 
uncertainties around the ICM viscosity and produce a prediction 
of the magnetic-field scale to be compared with the observed 
values reviewed in \secref{sec_clusters}. 

In the solar wind, the plasma is magnetised ($\rho_i\sim10^2$ km), 
while collisions are virtually absent: 
the mean free path exceeds the distance from the Sun ($10^8$ km). 
Ion pressure (temperature) anisotropies with respect 
to the field direction were directly measured in 1970s 
\cite{Feldman_etal_aniso,Marsch_Ao_Tu}. 
As was first suggested by Parker \cite{Parker_firehose}, 
firehose and mirror instabilities (as well as several others) 
should play a major, although not entirely understood 
role \cite{Gary_book}. A vast geophysical 
literature now exists on this subject, which cannot be reviewed here. 

\subsection{Kinetic turbulence} 
\label{sec_GK}

The instabilities are quenched when the magnetic field is sufficiently 
strong: $B^2$ overwhelms $\pperp-\ppar$ in the second term on the right-hand 
side of \eeqref{u_eq_brag}. 
If we use the collisional estimate \exref{Brag_visc}, this happens 
when 
\bea
B^2\gg \nupar\du_{\ld}/\ld\sim\Rey^{-1/2}\du_L^2. 
\label{quench}
\eea

The firehose-unstable perturbations become Alfv\'en waves in this limit. 
In the strong-field regime ($\dB\ll B_0$), 
the appropriate mathematical description of the 
weakly collisional turbulence of Alfv\'en waves 
is the low-frequency limit of the plasma kinetic theory called the 
gyrokinetics \cite{Frieman_Chen,SCDHHQ_gk}.\footnote{While 
originally developed and widely used for fusion plasmas, 
this ``kinetic-fluid'' description 
has only recently started to be applied to astrophysical 
problems such as the relative heating of ions and electrons 
by Alfv\'enic turbulence in advection-dominated 
accretion flows \cite{Quataert_Gruzinov_heating}.} 
It is obtained under an ordering scheme that stipulates 
\bea
\kperp\rho_i\sim1,\qquad
\omega/\Omega_i\sim\kpar/\kperp\sim\du/\uA\sim\dB/B_0\ll1. 
\label{GK_ordering}
\eea
The second relation in \eeqref{GK_ordering} coincides with 
the GS critical-balance conjecture \exref{crit_bal} 
if the latter is treated as an ordering assumption. 
The gyrokinetics can be cast as a systematic expansion 
of the full kinetic description of the plasma 
in the small parameter $\epsilon\sim\kpar/\kperp$ 
--- the direct generalisation of the similar expansion of 
MHD equations given at the end of \secref{sec_GS}. It turns out 
that the decoupling of the Alfv\'en-wave cascade that we demonstrated 
there is also a property of the gyrokinetics and that this 
cascade is correctly described by the RMHD equations \exsandref{RMHD_psi}{RMHD_phi} 
all the way down to the ion gyroscale, $\kperp\rho_i\sim1$ 
\cite{SCDHHQ_gk}. Broad fluctuation spectra observed in the solar 
wind \cite{Goldstein_Roberts} and in the ISM \cite{Armstrong_Rickett_Spangler} 
are likely to be manifestations of just such a cascade. 
The slow waves and the entropy mode 
are passively mixed by the Alfv\'en-wave cascade, but 
\eqsref{eq_drho}{eq_upar} have to be replaced by a kinetic 
equation. 

When magnetic fields are not stronger than 
the turbulent motions --- as is the case for clusters, 
where the magnetic energy is, in fact, quite close 
to the threshold \exref{quench} --- 
the situation is more complicated and much more obscure 
because small-scale dynamo (\secref{sec_dynamo}), 
back reaction (\secref{sec_sat}), plasma instabilities 
(in weak-field regions such as, for example, the bending 
regions of the folded fields, \secref{sec_dynamo}), and 
Alfv\'en waves (possibly of the kind discussed in \secref{sec_sat}) 
all enter into the mix and remain to be sorted out.

\section{Conclusion}

We conclude here in the hope that we have provided the reader 
with a fair overview of the state of affairs to which the MHD turbulence 
theory has arrived after its first fifty years. 
Perhaps, despite much insight gained along the way, not very far.  
It is clear that a simple extension of Kolmogorov's theory has so far 
proven unattainable. Two of the key assumptions of that theory --- 
{\bf isotropy} and {\bf locality of interactions} --- are manifestly 
incorrect for MHD. Indeed, even the applicability of the fundamental 
principle of small-scale universality is suspect. 
Although a fair amount 
is known about Alfv\'enic turbulence, progress in answering 
many important astrophysical questions (see the Introduction) 
has been elusive because there is 
little knowledge of the general spectral and structural properties 
of the fully developed turbulence in an MHD fluid and, more generally, 
in magnetised weakly collisional or collisionless plasmas. 
Thus, while many unanswered questions demand further effort on  
MHD turbulence, there is also an imperative, mandated by astrophysical 
applications, to go beyond the fluid description. 

\subparagraph{Acknowledgments.} 
We would like to thank Russell Kulsrud who is originally 
responsible for our interest in these matters. 
We are grateful to N.\ A.\ Lipunova, M.\ Bernard, 
R.\ A.\ Sunyaev, R.\ Buchstab and H.\ Lindsay 
for helping us obtain the photos used in this chapter. 
Our work was supported by the UKAFF Fellowship, 
PPARC Advanced Fellowship, King's College, Cambridge (A.A.S.) 
and in part by the US DOE Center for Multiscale Plasma Dynamics (S.C.C.). 

%\bibliography{sc_MHDbook}

\begin{thebibliography}{99.}

%\bibitem{Armstrong_Cordes_Rickett}
%Armstrong JW, Cordes JM, Rickett BJ (1981) 
%Density power spectrum in the local interstellar medium.
%Nature 291:561--564

\bibitem{Armstrong_Rickett_Spangler}
Armstrong JW, Rickett BJ, Spangler SR (1995) 
Electron density power spectrum in the local interstellar medium.
Astrophys J 443:209--221

\bibitem{Batchelor}
Batchelor GK (1950) 
On the spontaneous magnetic field in a conducting liquid in turbulent motion. 
Proc Roy Soc London A 201:405--416

\bibitem{Belcher_Davis}
Belcher JW, Davis L (1971)
Large-amplitude Alfv\'en waves in interplanetary medium, 2.
J Geophys Res 76:3534--3563

%\bibitem{Bhattacharjee_Ng01}
%Bhattacharjee A, Ng CS (2001)
%Random scattering and anisotropic turbulence of shear Alfv\'en wave packets. 
%Astrophys J 548:318--322

%\bibitem{Biermann_Schlueter}
%Biermann L, Schl\"uter A (1951) 
%Cosmic radiation and cosmic magnetic fields. 
%II. Origin of cosmic magnetic fields.
%Phys Rev 82:863--868

\bibitem{Biskamp_Mueller}
Biskamp D, M\"uller W-C (2000)
Scaling properties of three-dimensional isotropic magnetohydrodynamic turbulence. 
Phys Plasmas 7:4889--4900

\bibitem{Braginskii}
Braginskii SI (1965) 
Transport processes in a plasma. 
Rev Plasma Phys 1:205--310

\bibitem{Chandrasekhar_Kaufman_Watson}
Chandrasekhar S, Kaufman AN, Watson KM (1958) 
The stability of the pinch. 
Proc R Soc London A 245:435--455

\bibitem{Chertkov_etal_dynamo}
Chertkov M, Falkovich G, Kolokolov I, Vergassola M (1999)
Small-scale turbulent dynamo. 
Phys Rev Lett 83:4065--4068

\bibitem{Childress_Gilbert}
Childress S, Gilbert AD (1995) 
Stretch, Twist, Fold: The Fast Dynamo. Springer 

\bibitem{Cho_Lazarian_Vishniac}
Cho J, Lazarian A, Vishniac ET (2002)
Simulations of magnetohydrodynamic turbulence in a strongly magnetized medium. 
Astrophys J 564:291--301

\bibitem{Clarke_Kronberg_Boehringer}
Clarke TE, Kronberg PP, B\"ohringer H (2001)
A new radio-X-ray probe of galaxy cluster magnetic fields.
Astrophys J 547:L111--L114

\bibitem{Coleman}
Coleman PJ (1968)
Turbulence, viscosity, and dissipation in the solar-wind plasma. 
Astrophys J 153:371--388

\bibitem{Dennis_Chandran}
Dennis TJ, Chandran BDG (2005)
Turbulent heating of galaxy-cluster plasmas.
Astrophys J 622:205--216

\bibitem{Dobrowolny_Mangeney_Veltri}
Dobrowolny M, Mangeney A, Veltri P (1980)
Fully developed anisotropic hydromagnetic turbulence in the interplanetary space. 
Phys Rev Lett 45:144--147

\bibitem{Elsasser}
Elsasser WM (1950) The hydromagnetic equations. 
Phys Rev 79:183

%\bibitem{Fabian_etal_Perseus1}
%Fabian AC, Sanders JS, Allen SW, Crawford CS, Iwasawa K, Johnstone RM, 
%Schmidt RW, Taylor GB (2003)
%A deep {\em Chandra} observation of the Perseus cluster: shocks and ripples. 
%Mon Not R Astr Soc 344:L43--L47

\bibitem{Fabian_etal_Perseus2}
Fabian AC, Sanders JS, Crawford CS, Conselice CJ, Gallagher JS, Wyse RFG (2003)
The relationship between optical H$\alpha$ filaments and the X-ray emission in 
the core of the Perseus cluster.
Mon Not R Astr Soc 344:L48--L52

%\bibitem{Fabian_etal_Centaurus}
%Fabian AC, Sanders JS, Taylor GB, Allen SW (2005) 
%A deep {\em Chandra} observation of the Centaurus cluster: bubbles, filaments and edges.
%Mon Not R Astr Soc 360:L20-- \REM{Reference?}

\bibitem{Falkovich_Gawedzki_Vergassola}
Falkovich G, Gaw\c{e}dzki K, Vergassola M (2001)
Particles and fields in fluid turbulence. 
Rev Mod Phys 73:913--975

\bibitem{Feldman_etal_aniso}
Feldman WC, Asbridge JR, Bame SJ, Montgomery MD (1974)
Interpenetrating solar wind streams. 
Rev Geophys Space Phys 12:715--723 

\bibitem{Frieman_Chen}
Frieman EA, Chen L (1982)
Nonlinear gyrokinetic equations for low-frequency electromagnetic waves 
in general plasma equilibria. 
Phys Fluids 25:502--508

\bibitem{Gary_book}
Gary SP (1993)
Theory of space plasma microinstabilities. 
Cambridge: Cambridge University Press 

%\bibitem{Gary_etal}
%Gary SP, Mongomery MD, Feldman WC, Forslund DW (1976)
%Proton temperature instabilities in the solar wind. 
%J Geophys Res 81:1241--1246 

\bibitem{Galtier_etal}
Galtier S, Nazarenko SV, Newell AC, Pouquet A (2000) 
A weak turbulence theory for incompressible magnetohydrodynamics. 
J Plasma Phys 63:447--488

%\bibitem{Galtier_etal02}
%Galtier S, Nazarenko SV, Newell AC, Pouquet A (2002) 
%Anisotropic turbulence of shear-Alfv\'en waves.
%Astrophys J 564:L49--L52

\bibitem{Goldhirsch_Sulem_Orszag}
Goldhirsch I, Sulem P-L, Orszag SA (1987)
Stability and Lyapunov stability of dynamical systems: a differential approach 
and a numerical method. 
Physica D 27:311--337 

\bibitem{GS95}
Goldreich P, Sridhar S (1995)
Toward a theory of interstellar turbulence. 
II. Strong Alfv\'enic turbulence. 
Astrophys J 438:763--775

\bibitem{GS97}
Goldreich P, Sridhar S (1997)
Magnetohydrodynamic turbulence revisited. 
Astrophys J 485:680--688

\bibitem{Goldstein_Roberts}
Goldstein ML, Roberts DA (1999)
Magnetohydrodynamic turbulence in the solar wind.
Phys Plasmas 6:4154--4160 

\bibitem{Govoni_Feretti_review}
Govoni F, Feretti L (2004)
Magnetic fields in clusters of galaxies.
Int J Mod Phys D 13:1549--1594

\bibitem{Gruzinov_Diamond96}
Gruzinov AV, Diamond PH (1996)
Nonlinear mean field electrodynamics of turbulent dynamos. 
Phys Plasmas 3:1853--1857

\bibitem{Haugen_Brandenburg_Dobler}
Haugen NEL, Brandenburg A, Dobler W (2004)
Simulations of nonhelical hydromagnetic turbulence. 
Phys Rev E 70:016308

\bibitem{Higdon1}
Higdon JC (1984)
Density fluctuations in the interstellar medium: evidence for anisotropic 
magnetogasdynamic turbulence. I. Model and astrophysical sites.
Astrophys J 285:109--123 

%\bibitem{Higdon2}
%Higdon JC (1986)
%Density fluctuations in the interstellar medium: evidence for anisotropic 
%magnetogasdynamic turbulence. II. Stationary structures. 
%Astrophys J 309:342--361

\bibitem{Inogamov_Sunyaev}
Inogamov NA, Sunyaev RA (2003) 
Turbulence in clusters of galaxies and X-ray-line profiles. 
Astron Lett 29:791--824

\bibitem{Iroshnikov}
Iroshnikov RS (1963) 
Turbulence of a conducting fluid in a strong magnetic field. 
Astron Zh 40:742-750 [English translation: Sov Astron 7:566--571 (1964) 
Note that Iroshnikov's initials are given incorrectly as PS in the 
translation]

\bibitem{Kazantsev}
Kazantsev AP (1967) 
Enhancement of a magnetic field by a conducting fluid.
Zh Eksp Teor Fiz 53:1806--1813 
[English translation: Sov Phys JETP 26:1031--1034 (1968)]

\bibitem{Kim_etal_Coma}
Kim K-T, Kronberg PP, Dewdney PE, Landecker TL (1990)
The halo and magnetic field of the Coma cluster of galaxies. 
Astrophys J 355:29-37

\bibitem{K41}
Kolmogorov AN (1941)
The local structure of turbulence in incompressible viscous fluid at very 
large Reynolds numbers. 
Dokl Akad Nauk SSSR 30:299--303
[English translation: Proc Roy Soc London A 434:9--13 (1991)]

\bibitem{Kraichnan}
Kraichnan RH (1965) 
Inertial-range spectrum of hydromagnetic turbulence. 
Phys Fluids 8:1385--1387

\bibitem{Kraichnan_model}
Kraichnan RH (1968)
Small-scale structure of a scalar field convected by turbulence.
Phys Fluids 11:945--953

\bibitem{Kulsrud_Anderson}
Kulsrud RM, Anderson SW (1992)
The spectrum of random magnetic fields in the mean field dynamo theory 
of the galactic magnetic field. 
Astrophys J 396:606--630 

\bibitem{Landau_Lifshitz}
Landau LD, Lifshitz EM (1987)
Fluid Mechanics. %2nd Edition. 
Butterworth-Heinemann 

\bibitem{Lithwick_Goldreich01}
Lithwick Y, Goldreich P (2001)
Compressible magnetohydrodynamic turbulence in interstellar plasmas.
Astrophys J 562:279--296

\bibitem{Lithwick_Goldreich03}
Lithwick Y, Goldreich P (2003)
Imbalanced weak magnetohydrodynamic turbulence. 
Astrophys J 582:1220--1240

\bibitem{Maron_Goldreich}
Maron J, Goldreich P (2001)
Simulations of incompressible magnetohydrodynamic turbulence.
Astrophys J 554:1175--1196

%\bibitem{Marsch_review}
%Marsch E (1991)
%Turbulence in the solar wind. 
%Rev Mod Astron 4:145--156

\bibitem{Marsch_Ao_Tu}
Marsch E, Ao X-Z, Tu C-Y (2004)
On the temperature anisotropy of the core part of the proton velocity 
distribution function in the solar wind. 
J Geophys Res 109:A04102

%\bibitem{Marsch_etal_aniso}
%Marsch E, M\"uhlh\'auser K-H, Schwenn A, Rosenbauer H, Pilipp W, Neubauer FM (1982)
%Solar wind protons: three-dimensional velocity distributions and derived 
%plasma parameters measured between 0.3 and 1 AU. 
%J Geophys Res 87, 52 

\bibitem{Matthaeus_Goldstein}
Matthaeus WH, Goldstein ML (1982)
Measurement of the rugged invariants of magnetohydrodynamic turbulence in the solar wind.
J Geophys Res 87:6011--1028

\bibitem{Matthaeus_Goldstein_Roberts}
Matthaeus WH, Goldstein ML, Roberts DA (1990)
Evidence for the presence of quasi-two-dimensional nearly incompressible fluctuations in the solar wind.
J Geophys Res 95:20673--20683

\bibitem{Meneguzzi_Frisch_Pouquet}
Meneguzzi M, Frisch U, Pouquet A (1981)
Helical and nonhelical turbulent dynamos. 
Phys Rev Lett 47:1060--1064

\bibitem{Montgomery_Turner}
Montgomery D, Turner L (1981)
Anisotropic magnetohydrodynamic turbulence in a strong external magnetic field. 
Phys Fluids 24:825--831

\bibitem{Mueller_Biskamp_Grappin}
M\"uller W-C, Biskamp D, Grappin R (2003)
Statistical anisotropy of magntohydrodynamic turbulence.
Phys Rev E 67:066302

%\bibitem{Nazarenko_Newell_Galtier}
%Nazarenko SV, Newell AC, Galtier S (2001)
%Non-local MHD turbulence. 
%Physica D 152:646--652

\bibitem{Ng_Bhattacharjee96}
Ng CS, Bhattacharjee A (1996)
Interaction of shear-Alfv\'en wave packets: implication for weak magnetohydrodynamic
turbulence in astrophysical plasmas. 
Astrophys J 465:845--854

%\bibitem{Ng_Bhattacharjee97}
%Ng CS, Bhattacharjee A (1997)
%Scaling of anisotropic spectra due to the weak interaction of shear-Alfv\'en 
%wave packets. 
%Phys Plasmas 4:605--610

\bibitem{Obukhov}
Obukhov AM (1949) 
Structure of a temperature field in a turbulent flow.
Izv Akad Nauk SSSR Ser Geogr Geofiz 13:58--69

\bibitem{Ott_review}
Ott E (1998)
Chaotic flows and kinematic magnetic dynamos: A tutorial review. 
Phys Plasmas 5:1636--1646

\bibitem{Parker_solar_wind}
Parker EN (1958) 
Dynamics of the interplanetary gas and magnetic fields. 
Astrophys J 128:664--676 

\bibitem{Parker_firehose}
Parker EN (1958)
Dynamical instability in an anisotropic ionized gas of low density. 
Phys Rev 109:1874--1876

\bibitem{Quataert_Gruzinov_heating}
Quataert E, Gruzinov A (1999) 
Turbulence and particle heating in advection-dominated accretion flows. 
Astrophys J 520:248--255 

\bibitem{Rebusco_etal}
Rebusco P, Churazov E, B\"ohringer H, Forman W (2005)
Impact of stochastic gas motions on galaxy cluster abundance profiles.
Mon Not R Astr Soc 359:1041--1048

%\bibitem{Robinson_Rusbridge}
%Robinson DC, Rusbridge MG (1971)
%Structure of turbulence in the zeta plasma. 
%Phys Fluids 14:2499--2511

\bibitem{Rosenbluth}
Rosenbluth MN (1956)
Stability of the pinch. 
Los Alamos Scientific Laboratory Report LA-2030 

\bibitem{SCMM_folding}
Schekochihin A, Cowley S, Maron J, Malyshkin L (2002)
Structure of small-scale magnetic fields in the kinematic dynamo theory.
Phys Rev E 65:016305 

\bibitem{SCHMM_ssim}
Schekochihin AA, Cowley CS, Hammett GW, Maron JL, McWilliams JC (2002) 
A model of nonlinear evolution and saturation of the turbulent MHD dynamo.
New J Phys 4:84

%\bibitem{SCTHMM_aniso}
%Schekochihin AA, Cowley CS, Taylor SF, Hammett GW, Maron JL, McWilliams JC (2004) 
%Saturated state of the nonlinear small-scale dynamo.
%Phys Rev Lett 92:084504

\bibitem{SCTMM_stokes}  
Schekochihin AA, Cowley CS, Taylor SF, Maron JL, McWilliams JC (2004) 
Simulations of the small-scale turbulent dynamo. 
Astrophys J 612:276--307

\bibitem{SCKHS_brag}  
Schekochihin AA, Cowley CS, Kulsrud RM, Hammett GW, Sharma P (2005) 
Plasma instabilities and magnetic-field growth in clusters of galaxies. 
Astrophys J 629:139--142 

\bibitem{SCDHHQ_gk}  
Schekochihin AA, Cowley CS, Dorland WD, Hammett GW, Howes GG, Quataert E, 
Tatsuno T (2007) 
Kinetic and fluid turbulent cascades in magnetized weakly collisional 
astrophysical plasmas. Astrophys J (submitted; E-print {\tt arXiv:0704.0044})

\bibitem{Schlueter_Biermann}
Schl\"uter A, Biermann L (1950)
Interstellare Magnetfelder. 
Z Naturforsch 5a:237--251

\bibitem{Schuecker_etal}
Schuecker P, Finoguenov A, Miniati F, B\"ohringer H, Briel UG (2004)
Probing turbulence in the Coma galaxy cluster. 
Astron Astrophys 426:387--397

\bibitem{Shakura_Sunyaev}
Shakura NI, Sunyaev RA (1973)
Black holes in binary systems. Observational appearance. 
Astron Astrophys 24:337--355

\bibitem{Shebalin_Matthaeus_Montgomery}
Shebalin JV, Matthaeus WH, Montgomery D (1983)
Anisotropy in MHD turbulence due to a mean magnetic field. 
J Plasma Phys 29:525--547

\bibitem{Spitzer_book}
Spitzer L (1962)
Physics of fully ionized gases. 
New York: Wiley 

%\bibitem{Spitzer_Haerm}
%Spitzer L, H\"arm R (1952)
%Transport phenomena in a completely ionized gas. 
%Phys Rev 89:977--981

\bibitem{Sridhar_Goldreich}
Sridhar S, Goldreich P (1994) 
Toward a theory of interstellar turbulence. 
I. Weak Alfv\'enic turbulence.  
Astrophys J 432:612--621

\bibitem{Strauss}
Strauss HR (1976) 
Nonlinear, three-dimensional magnetohydrodynamics of noncircular tokamaks.
Phys Fluids 19:134--140

\bibitem{Vedenov_Sagdeev}
Vedenov AA, Sagdeev RZ (1958) 
Some properties of the plasma with anisotropic distribution of the velocities 
of ions in the magnetic field. 
Sov Phys Dokl 3:278

%\bibitem{Vogt_Ensslin1}
%Vogt C, En{\ss}lin TA (2003)
%Measuring the cluster magnetic field power spectra from Faraday rotation 
%maps of Abell 400, Abell 2634 and Hydra A.
%Astron Astrophys 412:373--385

\bibitem{Vogt_Ensslin2}
Vogt C, En{\ss}lin TA (2005)
A Bayesian view on Fraday rotation maps --- Seeing the magnetic power spectra 
in galaxy clusters. 
Astron Astrophys 434:67--76

\bibitem{Willson}
Willson MAG (1970) 
Radio observations of the cluster of galaxies in Coma Berenices --- The 5C4 survey.
Mon Not R Astr Soc 151:1--44

\bibitem{Zeldovich_antidynamo}
Zeldovich YaB (1956)
The magnetic field in the two-dimensional motion of a conducting 
turbulent liquid. 
Zh Exp Teor Fiz 31:154--155
[English translation: Sov Phys JETP 4:460--462 (1957)]

\bibitem{Zeldovich_etal_linear}
Zeldovich YaB, Ruzmaikin AA, Molchanov SA, Sokoloff DD (1984)
Kinematic dynamo problem in a linear velocity field.
J Fluid Mech 144:1--11

\bibitem{Almighty_Chance}
Zeldovich YaB, Ruzmaikin AA, Sokoloff DD (1990) 
The Almighty Chance. Singapore: World Scientific 

%\bibitem{Zweben_Menyuk_Taylor}
%Zweben SJ, Menyuk CR, Taylor RJ (1979)
%Small-scale magnetic fluctuations inside the Macrotor tokamak. 
%Phys Rev Lett 42:1270--1274

\end{thebibliography}
%\end{document}

\printindex

\end{document}